\documentclass[aps,pra,10pt,superscriptaddress,twocolumn,nofootinbib,floatfix]{revtex4-2}
\usepackage{fouriernc}
\usepackage{graphicx}
\usepackage{amsmath,amsthm,amssymb,dsfont,soul}
\usepackage{mdframed}
\usepackage{orcidlink}
\usepackage{subcaption}
\newcommand{\ket}[1]{\left| #1 \right\rangle}
\newcommand{\bra}[1]{\left\langle #1 \right|}

\newcommand{\braket}[2]{\langle #1|#2 \rangle}
\newcommand{\ketbra}[2]{\left|#1\right\rangle\hskip-1mm\left\langle#2\right|}

\usepackage[capitalise]{cleveref}
\crefname{equation}{Eq.}{Eqs.}
\crefname{figure}{Fig.}{Figs.}
\usepackage[authormarkuptext=name,commentmarkup=uwave]{changes}
\definechangesauthor[name={JRH},color=green!70!black]{jonte}

\definechangesauthor[name={NGU},color=Magenta!70!black]{nick}

\usepackage{epstopdf}
\usepackage[dvipsnames]{xcolor}
\usepackage{bm}
\usepackage{pstool}
\usepackage[percent]{overpic}
\usepackage{rotating}
\usepackage[normalem]{ulem}
\usepackage{array}

\usepackage{physics}
\usepackage{tikz}
\usepackage{mathdots}
\usepackage{cancel}
\usepackage{multirow}
\usepackage{tabularx}
\usepackage{extarrows}
\usepackage{booktabs}
\usetikzlibrary{fadings}
\usetikzlibrary{patterns}
\usetikzlibrary{shadows.blur}
\usetikzlibrary{shapes}

\usepackage{mathrsfs}
\usepackage{xfrac}
\usepackage{enumitem}
\usepackage{mathtools}
\usepackage{nccmath}
\usepackage{algorithm}
\usepackage[noEnd=false]{algpseudocodex}

\setlist[itemize]{noitemsep, nolistsep}

\definecolor{Xred}{HTML}{B31B1B}

\definecolor{verylightgray}{HTML}{F3F3F3}
\definecolor{notsoverylightgray}{HTML}{E3E3E3}

\newtheorem*{theorem*}{Theorem}
\newtheorem*{corollary*}{Corollary}
\newtheorem*{lemma*}{Lemma}

\newtheorem*{assumption*}{Assumption}

\newtheorem*{principle*}{Principle}

\newtheorem{strategy}{Strategy}
\newtheorem{procedure}{Procedure}

\DeclarePairedDelimiterX\projector[1]{\lvert}{\rvert}{#1\delimsize\rangle\!\delimsize\langle#1}

\def\Turn{{\color{Sepia}\texttt{Turn}}}
\def\Entropy{{\color{Sepia}\texttt{Entropy}}}
\def\Priors{{\color{Sepia}\texttt{Priors}}}
\def\ProbDgB{{\color{Sepia}\texttt{ProbDgB}}}
\def\ProbBgD{{\color{Sepia}\texttt{ProbBgD}}}
\def\PlayerGuess{{\color{Sepia}\texttt{PlayerGuess}}}
\def\InputState{{\color{Sepia}\texttt{InputState}}}
\def\TrialInputState{{\color{Sepia}\texttt{TrialInputState}}}
\def\ProbD{{\color{Sepia}\texttt{ProbD}}}
\def\PotEntropy{{\color{Sepia}\texttt{PotEntropy}}}
\def\ExpectedEntropy{{\color{Sepia}\texttt{ExpectedEntropy}}}
\def\ExpectedInfoGain{{\color{Sepia}\texttt{ExpectedInfoGain}}}

\begin{document}

\title{The Quantum Plumber's Problem}

\author{Nicolas G. Underwood\,\orcidlink{0000-0003-4803-2629}}
\email{nick.underwood@newcastle.ac.uk}
\affiliation{Quantum Group, School of Computing, Newcastle University, 1 Science Square, Newcastle upon Tyne, NE4 5TG, UK}

\author{Holger F. Hofmann}
\affiliation{Graduate School of Advanced Science and Engineering, Hiroshima University, Kagamiyama 1-3-1, Higashi Hiroshima 739-8530, Japan}

\author{Jonte R. Hance\,\orcidlink{0000-0001-8587-7618}}
\email{jonte.hance@newcastle.ac.uk}
\affiliation{Quantum Group, School of Computing, Newcastle University, 1 Science Square, Newcastle upon Tyne, NE4 5TG, UK}

\begin{abstract}
Recent work quantified the notion of quantum counterfactual gain for an extended Elitzur-Vaidman bomb test style scenario, through a connection to the negativity of the Kirkwood-Dirac quasiprobability distribution.
We here extend this work to identifying quantum advantage in a new scenario, which we term the ``Quantum Plumber's Problem''.
In this scenario, we imagine a ``quantum plumber'', who knows that one path of an interferometer is blocked, and wants to find the optimal strategy for identifying with certainty \emph{which} path this is.
We discuss various strategies for a generalised path-encoded interferometer, as well as for the specific case of Hofmann's three-path interferometer, introduced in a recent analysis of the relationship between states in five measurement contexts of a three level system. 
We support our arguments on the relative merit of competing strategies with data collected over many simulated attempts at locating blockages.
We also present results for a variant of the game in which the blockage is replaced by a non-demolition detector.
\end{abstract}

\maketitle

\section{Introduction}

Two of the authors recently quantified the notion of quantum counterfactual gain---the advantage quantum interaction-free measurement gives over classical un-interfering particle-type strategies---for identifying the presence or absence of an absorber in a given path of an interferometer~\cite{hance_counterfactuality_2024}.
Our definition for this gain was associated with the presence of extra terms appearing in the quantum version of the total variation distance between conditional detector activation probabilities in the presence and absence of a blockage. 
These in turn were associated with both a traditional sequential probability term (which we call the Elitzur-Vaidman term), \cite{elitzur_quantum_1993} and a Kirkwood-Dirac quasiprobability term \cite{gherardiniQuasiprobabilitiesQuantumThermodynamics2024,ArvidssonShukur2024KD,Kirkwood1933KD,Dirac1945KDDistr}.
Counterfactual gain, we argued, could be maximised through obtaining the maximum negative value of the latter of these two terms - a scenario no-one had yet considered. 
While extending the traditional Elitzur-Vaidman scheme, the scenario we presented however could still be considered a binary discrimination task between the absence or presence of a blockage in a predefined location, which previous work claims to have fully solved~\mbox{\cite{Rudolph2000Better}}. 

Here we instead extend this work to identifying quantum advantage in a new $|\mathcal{B}|$-ary discrimination scenario, which we term the ``Quantum Plumber's Problem''. In this scenario, we imagine a ``quantum plumber'', who knows that a single path of an interferometer is blocked, and wants to find \emph{which} path this is.
To do this, the plumber (who knows the architecture of the interferometer: i.e., the reflectivities of each beamsplitter and any phases applied by each path) is able to input single photons, in whatever superposition of the input ports they wish, and observe which output ports (if any) they emerge from.
Beyond this, the plumber is not allowed to change the interferometer at all. 
For the interferometers in question, there is generally substantially more locations in which the blockage may be found than output ports, and so the plumber requires multiple photons, which are entered sequentially, so that they may learn from the outcome of each and adapt their next photon accordingly. 
As detector outcomes are inherently probabilistic, the plumber must focus more on overall strategy than individual good turns; a plumber following a deterministically defined strategy will rarely use the same set of photons in two consecutive games.

We again present two rulesets for this scenario: one where the photon is treated ``classically''---i.e., as a non-interfering single-particle, like a pinball or pachinko ball, reflected or transmitted probabilistically at each beamsplitter based on the beamsplitter's reflectivity; and one where the photon is a standard quantum-mechanical photon, able to interfere with itself at beamsplitters as normal.
For the classical ruleset, we know from Ref.~\cite{hance_counterfactuality_2024} that a blocked path can only reduce the probability of the photon arriving at each detector.
However, in this game one path is always blocked, and the plumber must compare the effect of the blocker being at each possible location.
This means we can't simply associate increased detector activation probability with ``quantumness'' as we did in Ref.~\cite{hance_counterfactuality_2024}. Instead we need a more nuanced strategy to identify the advantage quantumness brings us.
To do this, we treat the scenario as a game, for which we present an analysis of multiple competing strategies, applicable (in varying degrees) to both the classical and the quantum ruleset. We supplement our analytic treatment of a generalized interferometer (\Cref{fig:interferometer}) with data obtained over many simulated games for a specific ``game-board'': Hofmann's three-path interferometer (\Cref{fig:Hofmann_interferometer}), introduced in a recent study of the measurement contextuality in five contexts of a three-level system~\cite{Hofmann2025Sphere}.
While the conclusions of our numerical comparison do not necessarily carry over to other instantiations of the game, they allow players to gain an intuition for how the game works and why certain strategies function more effectively, so may help guide future development of strategies. 

The physical problem is roughly analogous to that of quantum channel discrimination (QCD) \cite{chiribellaMemoryEffectsQuantum2008,duanPerfectDistinguishabilityQuantum2009,pianiAllEntangledStates2009,Harrow2010AdaptiveChannel,Pirandola2019ChannelDiscrimination,Zhuang2020ChannelPositionFinding,fangAdversarialQuantumChannel2025}.
A generalised QCD problem may attempt to discriminate between two or more channels, by introducing a choice of state preparations, choices of POVM measurements, multiple evaluations/queries with or without adaptive strategies, and the possibility of an entangled ancillary system.
Here the plumber is restricted to the same projective measurement on each turn (dictated by the interferometer geometry), and is not permitted an ancillary system. 
The only freedom they are permitted is in the state preparation that they may adapt on each turn.
A key difference to QCD is that the blockage is trace reducing, and so not a channel.
As we explain, it is always possible to associate a blockage $\beta$ to a state $\ket{\beta}$, and as long as these are all distinct, it is possible to distinguish between blockage locations to find the blockage with certainty in a finite number of turns.

Locating an absorber among several paths has been considered before, both by chaining two-path interaction-free measurements together~\mbox{\cite{Filatov2024MultipleIFM,Franco2026MultipleIFM}} and, in recent work by Chaturvedi \emph{et al}~\mbox{\cite{Chaturvedi2026IFLocalization}} for a multiport setup.
In this latter study, the authors designed an optimal interferometer setup to interrogate multiple orthogonal blockage locations. 
Our plumber does not have this luxury: our possible blockage locations are not necessarily orthogonal, meaning that if the photon is absorbed, it is not possible to tell with certainty which of the potential $|\mathcal{B}|$ locations it was absorbed at.

This paper is laid out as follows. In \Cref{sec:Game}, we define the Quantum Plumber's Problem game, first giving an overview, then describing the interferometer setup and notation used, before explaining the core turn mechanics, and the instructions for the Gamemaster (GM), who adopts the role of the laws of physics. We then give the rules for calculating the (standard quantum) probabilities we would expect for photons, as well as an alternative ruleset for ``classical'' photons (acting like pin-balls, in a Pachinko machine or Galton board), to allow us to define an advantage given by quantum mechanics.

In \Cref{sec:GameplayAnalysis}, we give an initial analysis of gameplay. We first introduce two metrics for game progress, which we then use to build competing strategies around. 
We discuss a strategy centering around selecting input photon states with the aim of ruling out blockage locations.
We then construct an information theoretic strategy that aims to maximise information gain, as well as an important sampling pitfall that can prevent this strategy from completing games.
After briefly discussing mixed states, we compare the performance of variations on the introduced strategies, both with the classical and quantum rulesets, applied to Hofmann's three-path interferometer.

Finally, in \Cref{sec:Discussion}, we summarise both the game rules and the game strategy and analysis discussed so far, and discuss a variant of the game in which the blockage is a non-demolition detector, before explaining avenues for potential future works: extensions of the game, additional strategies, and ways in which the game or strategies for it link to myriad different areas within quantum information theory, game theory, expected utility theory, and machine learning.

\section{The Game}\label{sec:Game}
\subsection{Game overview}
The premise of the game is for the player to take the role of a ``Quantum Plumber'', tasked with fixing a broken interferometer. The interferometer in question contains a single blocked path, which the player must attempt to locate. The interferometer is however sealed in a black box, so the player must attempt to locate the blockage by sending photons into the interferometer's input ports in order to learn from the response of the single photon detectors coupled to each of the interferometer's output ports. To assist them, the player is presented with a schematic of the working interferometer (i.e., the interferometer architecture without any blockages) to enable them to calculate the probability $P(\omega|\beta)$ of each detector $\omega$ activating given a blockage at each of the potential blockage locations $\beta$.
Each photon sent into the interferometer is considered to take a player turn, and the game concludes when the blocked path has been located with unit probability. The goal of the game is to find the blockage in as few turns as possible. 

\subsection{Interferometer setup and notation}
\label{sec:interferometer}

\begin{figure}
\input{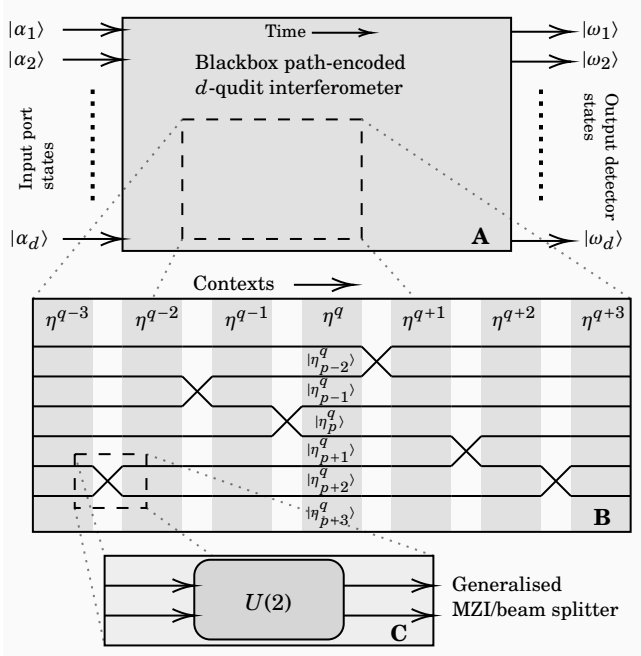}
    \caption{The path-encoded $d$-state qudit interferometer setup we consider, with A) The outline setup, with input ports labelled $\ket{\alpha_1}...\ket{\alpha_d}$ and output ports and corresponding photon detectors labelled $\ket{\omega_1}...\ket{\omega_d}$. B) The vertical cross-sections between nodes corresponding to differing complete orthonormal bases that we refer to as ``contexts''. C) Nodes containing a generalised MZI/beam splitter capable of performing any transformation in $U(2)$.}
    \label{fig:interferometer}
\end{figure}

For our interferometer, let us use a path-encoded $d$-state (qudit) optical system~\cite{erhard_advances_2020}.\footnote{We can of course build an optical interferometer using degrees of freedom other than path, e.g., by using polarisation or orbital angular momentum modes, alongside or instead of path; however, here we consider our only degree of freedom to be path, as is the case in most integrated photonics chips, which can only support a single polarisation mode.} These have long been known to be capable of realising any operator in $U(d)$ \cite{reck_experimental_1994}, and are currently employed as integral components in linear optical quantum computers \cite{carolan_universal_2015,wang_qudits_2020,paesani_scheme_2021,chi_programmable_2022}.
We represent the interferometer as in \Cref{fig:interferometer}, with a circuit diagram comprising $d$ main paths, with adjacent paths sequentially connected at nodes containing a generalised beam splitter or Mach-Zehnder interferometer (MZI) capable of performing any element of $U(2)$; an input port connected to the left-hand side of each path; and a photon detector connected to each of the $d$ output ports at the right-hand side of each path. 
We refer to the (orthonormal) state vectors corresponding to the input ports as $\ket{\alpha_1},...,\ket{\alpha_d}$, and those corresponding to the output ports as $\ket{\omega_1},...,\ket{\omega_d}$.
Together the photon detectors act to perform a traditional projective measurement, so that in the absence of a blockage the total probability of a detector firing $\sum_{i=1}^dP(\omega_i)$ is unity each time a photon is entered.\footnote{Note we consider perfect lossless components; for realistic lossy components this would necessarily be less than one. We discuss extension to this realistic scenario in \Cref{sec:Discussion}.}

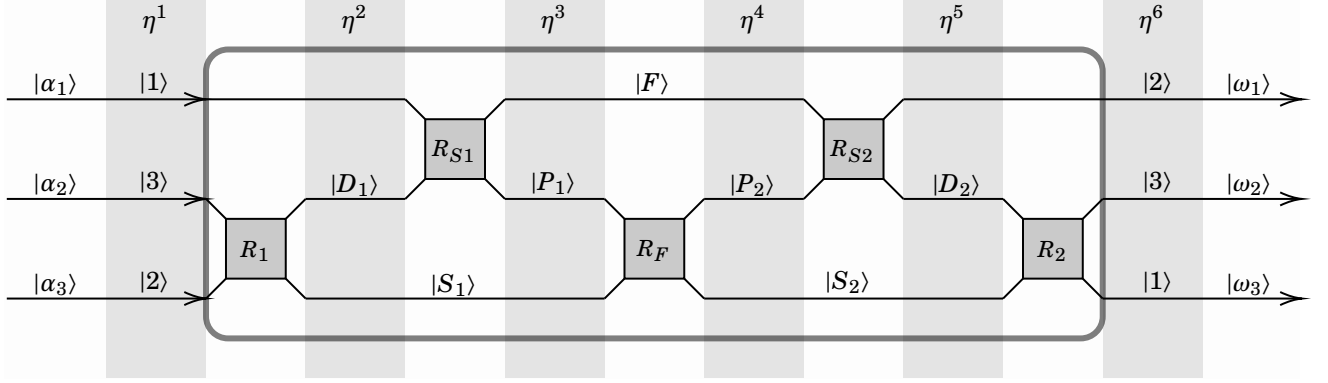
\begin{figure*}
\tikzset{every picture/.style={line width=0.75pt}} 

\begin{tikzpicture}[x=0.75pt,y=0.75pt,yscale=-1,xscale=1]

\draw  [draw opacity=0][fill={rgb, 255:red, 0; green, 0; blue, 0 }  ,fill opacity=0.01 ] (0,-25) -- (650,-25) -- (650,190) -- (0,190) -- cycle ;
\draw  [color={rgb, 255:red, 0; green, 0; blue, 0 }  ,draw opacity=0.5 ][line width=2.25]  (101,36) .. controls (101,29.92) and (105.92,25) .. (112,25) -- (540,25) .. controls (546.08,25) and (551,29.92) .. (551,36) -- (551,159) .. controls (551,165.08) and (546.08,170) .. (540,170) -- (112,170) .. controls (105.92,170) and (101,165.08) .. (101,159) -- cycle ;
\draw  [draw opacity=0][fill={rgb, 255:red, 0; green, 0; blue, 0 }  ,fill opacity=0.1 ] (151,0) -- (201,0) -- (201,190) -- (151,190) -- cycle ;
\draw  [draw opacity=0][fill={rgb, 255:red, 0; green, 0; blue, 0 }  ,fill opacity=0.1 ] (51,0) -- (101,0) -- (101,190) -- (51,190) -- cycle ;
\draw  [draw opacity=0][fill={rgb, 255:red, 0; green, 0; blue, 0 }  ,fill opacity=0.1 ] (251,0) -- (301,0) -- (301,190) -- (251,190) -- cycle ;
\draw  [draw opacity=0][fill={rgb, 255:red, 0; green, 0; blue, 0 }  ,fill opacity=0.1 ] (351,0) -- (401,0) -- (401,190) -- (351,190) -- cycle ;
\draw  [draw opacity=0][fill={rgb, 255:red, 0; green, 0; blue, 0 }  ,fill opacity=0.1 ] (451,0) -- (501,0) -- (501,190) -- (451,190) -- cycle ;
\draw  [draw opacity=0][fill={rgb, 255:red, 0; green, 0; blue, 0 }  ,fill opacity=0.1 ] (551,0) -- (601,0) -- (601,190) -- (551,190) -- cycle ;
\draw    (151,150) -- (301,150) ;
\draw    (201,50) -- (211,60) ;
\draw    (351,100) -- (401,100) ;
\draw    (351,150) -- (501,150) ;
\draw    (451,100) -- (501,100) ;
\draw    (151,100) -- (201,100) ;
\draw    (251,100) -- (301,100) ;
\draw    (251,50) -- (401,50) ;
\draw    (101,50) -- (201,50) ;
\draw    (201,100) -- (211,90) ;
\draw    (251,100) -- (241,90) ;
\draw  [fill={rgb, 255:red, 0; green, 0; blue, 0 }  ,fill opacity=0.2 ] (211,60) -- (241,60) -- (241,90) -- (211,90) -- cycle ;
\draw    (241,60) -- (251,50) ;
\draw    (101,100) -- (111,110) ;
\draw    (101,150) -- (111,140) ;
\draw    (151,150) -- (141,140) ;
\draw  [fill={rgb, 255:red, 0; green, 0; blue, 0 }  ,fill opacity=0.2 ] (111,110) -- (141,110) -- (141,140) -- (111,140) -- cycle ;
\draw    (141,110) -- (151,100) ;
\draw    (401,50) -- (411,60) ;
\draw    (401,100) -- (411,90) ;
\draw    (451,100) -- (441,90) ;
\draw  [fill={rgb, 255:red, 0; green, 0; blue, 0 }  ,fill opacity=0.2 ] (411,60) -- (441,60) -- (441,90) -- (411,90) -- cycle ;
\draw    (441,60) -- (451,50) ;
\draw    (501,100) -- (511,110) ;
\draw    (501,150) -- (511,140) ;
\draw    (551,150) -- (541,140) ;
\draw  [fill={rgb, 255:red, 0; green, 0; blue, 0 }  ,fill opacity=0.2 ] (511,110) -- (541,110) -- (541,140) -- (511,140) -- cycle ;
\draw    (541,110) -- (551,100) ;
\draw    (301,100) -- (311,110) ;
\draw    (301,150) -- (311,140) ;
\draw    (351,150) -- (341,140) ;
\draw  [fill={rgb, 255:red, 0; green, 0; blue, 0 }  ,fill opacity=0.2 ] (311,110) -- (341,110) -- (341,140) -- (311,140) -- cycle ;
\draw    (341,110) -- (351,100) ;
\draw    (1,50) -- (101,50) ;
\draw [shift={(101,50)}, rotate = 180] [color={rgb, 255:red, 0; green, 0; blue, 0 }  ][line width=0.75]    (10.93,-3.29) .. controls (6.95,-1.4) and (3.31,-0.3) .. (0,0) .. controls (3.31,0.3) and (6.95,1.4) .. (10.93,3.29)   ;
\draw    (1,100) -- (101,100) ;
\draw [shift={(101,100)}, rotate = 180] [color={rgb, 255:red, 0; green, 0; blue, 0 }  ][line width=0.75]    (10.93,-3.29) .. controls (6.95,-1.4) and (3.31,-0.3) .. (0,0) .. controls (3.31,0.3) and (6.95,1.4) .. (10.93,3.29)   ;
\draw    (1,150) -- (101,150) ;
\draw [shift={(101,150)}, rotate = 180] [color={rgb, 255:red, 0; green, 0; blue, 0 }  ][line width=0.75]    (10.93,-3.29) .. controls (6.95,-1.4) and (3.31,-0.3) .. (0,0) .. controls (3.31,0.3) and (6.95,1.4) .. (10.93,3.29)   ;
\draw    (451,50) -- (649,50) ;
\draw [shift={(651,50)}, rotate = 180] [color={rgb, 255:red, 0; green, 0; blue, 0 }  ][line width=0.75]    (10.93,-3.29) .. controls (6.95,-1.4) and (3.31,-0.3) .. (0,0) .. controls (3.31,0.3) and (6.95,1.4) .. (10.93,3.29)   ;
\draw    (551,100) -- (649,100) ;
\draw [shift={(651,100)}, rotate = 180] [color={rgb, 255:red, 0; green, 0; blue, 0 }  ][line width=0.75]    (10.93,-3.29) .. controls (6.95,-1.4) and (3.31,-0.3) .. (0,0) .. controls (3.31,0.3) and (6.95,1.4) .. (10.93,3.29)   ;
\draw    (551,150) -- (649,150) ;
\draw [shift={(651,150)}, rotate = 180] [color={rgb, 255:red, 0; green, 0; blue, 0 }  ][line width=0.75]    (10.93,-3.29) .. controls (6.95,-1.4) and (3.31,-0.3) .. (0,0) .. controls (3.31,0.3) and (6.95,1.4) .. (10.93,3.29)   ;

\draw (75.48,10) node  [font=\normalsize]  {$\eta ^{1}$};
\draw (175.63,93.19) node  [font=\normalsize]  {$|D_{1} \rangle $};
\draw (175.48,10) node  [font=\normalsize]  {$\eta ^{2}$};
\draw (275.48,10) node  [font=\normalsize]  {$\eta ^{3}$};
\draw (375.48,10) node  [font=\normalsize]  {$\eta ^{4}$};
\draw (475.48,10) node  [font=\normalsize]  {$\eta ^{5}$};
\draw (575.48,10) node  [font=\normalsize]  {$\eta ^{6}$};
\draw (225.24,143.19) node  [font=\normalsize]  {$|S_{1} \rangle $};
\draw (275.21,92.19) node  [font=\normalsize]  {$|P_{1} \rangle $};
\draw (325.01,42.2) node  [font=\normalsize]  {$|F\rangle $};
\draw (477.63,93.19) node  [font=\normalsize]  {$|D_{2} \rangle $};
\draw (410,134.4) node [anchor=north west][inner sep=0.75pt]  [font=\normalsize]  {$|S_{2} \rangle $};
\draw (375.21,93.19) node  [font=\normalsize]  {$|P_{2} \rangle $};
\draw (225.34,76.09) node  [font=\small]  {$R_{S1}$};
\draw (125.72,126.09) node  [font=\small]  {$R_{1}$};
\draw (425.34,76.09) node  [font=\small]  {$R_{S2}$};
\draw (525.72,126.09) node  [font=\small]  {$R_{2}$};
\draw (325.72,126.09) node  [font=\small]  {$R_{F}$};
\draw (25.42,42.81) node  [font=\normalsize]  {$| \alpha _{1} \rangle $};
\draw (75.03,41.8) node  [font=\normalsize]  {$|1 \rangle $};
\draw (25.42,92.81) node  [font=\normalsize]  {$| \alpha _{2} \rangle $};
\draw (75.03,91.8) node  [font=\normalsize]  {$|3 \rangle $};
\draw (25.42,142.81) node  [font=\normalsize]  {$| \alpha _{3} \rangle $};
\draw (75.03,141.8) node  [font=\normalsize]  {$|2 \rangle $};
\draw (624.45,42.81) node  [font=\normalsize]  {$| \omega _{1} \rangle $};
\draw (578.53,41.8) node  [font=\normalsize]  {$|2 \rangle $};
\draw (624.45,92.81) node  [font=\normalsize]  {$| \omega _{2} \rangle $};
\draw (578.53,91.8) node  [font=\normalsize]  {$|3 \rangle $};
\draw (624.45,142.81) node  [font=\normalsize]  {$| \omega _{3} \rangle $};
\draw (578.53,141.8) node  [font=\normalsize]  {$|1 \rangle $};

\end{tikzpicture}
    \caption{Hofmann's three-path interferometer \cite{hofmann_sequential_2023}. Here the beamsplitters have reflectivities $R_1=R_2=1/2$, $R_{S_1}=R_{S_2}=1/3$, and $R_F=1/4$, and apply a $\pi$ phase shift to reflections from/to the middle path. The labelled paths correspond to the states
    $\ket{D_1}=(\ket{2}-\ket{3})/\sqrt{2}$,
    $\ket{S_1}=(\ket{2}+\ket{3})/\sqrt{2}$,
    $\ket{P_1}=(2\ket{1}-\ket{2}+\ket{3})/\sqrt{6}$,
    $\ket{F}=(\ket{1}+\ket{2}-\ket{3})/\sqrt{3}$, 
    $\ket{P_2}=(-\ket{1}+2\ket{2}+\ket{3})/\sqrt{6}$,
    $\ket{S_2}=(\ket{1}+\ket{3})/\sqrt{2}$, and
    $\ket{D_2}=(\ket{1}-\ket{3})/\sqrt{2}$.}
    \label{fig:Hofmann_interferometer}
\end{figure*}

We assume there is a single blockage (no more, no fewer\footnote{See \Cref{sec:Discussion} for discussion of potential extensions to the game, including games where there are multiple blockers in play.}), and that it is located on a path segment connecting two nodes. We refer to these possible positions for the blockage as $\beta_1,...,\beta_{|\mathcal{B}|}$, where $\mathcal{B}$ is the set of such locations.
In the quantum formalism, each blockage location is associated with a state vector, $\ket{\beta_1},...,\ket{\beta_{|\mathcal{B}|}}$.
Beam splitters perform unitary transformations, so the paths before the beam splitter belong to a different orthogonal basis than the ones after the beam splitter. 
Following the terminology of Ref.~\cite{hofmann_sequential_2023}, we refer to each orthogonal basis as a ``context''. 
Since blockages can belong to different contexts, their state vectors are not necessarily orthogonal, adding a level of complexity to the problem that highlights the quantum nature of our scenario.

Given the comparative complexity of the general case of the Quantum Plumber's Problem over e.g., the blocking-or-not-blocking game considered in Ref.~\cite{hance_counterfactuality_2024}, we have supported our analytic arguments with data obtained numerically for Hofmann's three-path qutrit interferometer~\cite{hofmann_sequential_2023} (depicted in \Cref{fig:Hofmann_interferometer}) which was designed to investigate contextuality. Ignoring the trivial input/output paths, Hofmann's interferometer has seven potential blockage locations (i.e., $|\mathcal{B}|=7$).

\subsection{Core turn mechanics}
The player may treat the position of the blockage as a random variable $B$, which takes a value $\beta$ in the set of possible blockage locations $\mathcal{B}$. Similarly, the outcome of a turn (the detector response) is a random variable $\Omega$, which takes values $\omega$ in the set $\mathcal{O}=\{\omega_0,\omega_1,...,\omega_d\}$. We denote the event that no detector activates (i.e., for our perfect lossless interferometer that the photon is absorbed by the blockage) as $\omega_0$, and the event that the detector on the $p^\text{th}$ path activates as $\omega_p$. Unlike $B$, the random variable $\Omega$ depends on the input photon state, and so is influenced by the player's choice each turn. On each turn, the player is permitted to choose any pure photon input state $\ketbra{\psi}{\psi}$ where
\begin{equation}
\ket{\psi}=\sum_{p=1}^d a_p \ket{\alpha_p},\quad 
a_p\in\mathds{C},\quad
\sum_{p=1}^d|a_p|^2=1,
\end{equation}
or any input density matrix $\rho_\text{Qu}$, where
\begin{equation}\label{eq:density_matrix}
\begin{split}
\rho_\text{Qu}\in\mathds{C}^{d\times d},\quad
\rho_\text{Qu}^\dagger=\rho_\text{Qu},\\
\bra{\psi}\rho_\text{Qu}\ket{\psi}\geq0\,\,\,\forall\psi,\quad
\Tr[\rho_\text{Qu}]=1.
\end{split}
\end{equation}
(Note that in the Schr\"{o}dinger picture, the blockage acts to reduce the normalisation of $\ket{\psi}$ and trace of $\rho_\text{Qu}$.)
It is in the interest of the player to choose input states judiciously, so as to favourably influence random variable $\Omega$ in order to complete the game quickly.

Before any single turn, the player's knowledge of the blockage location is described by a set of prior probabilities, $P(\beta)$. At the start of the game these are initialised to $P(\beta)=1/|\mathcal{B}|$ for all $\beta$, to reflect total ignorance of the blockage location. On taking a turn, these priors $P(\beta)$ must be updated to reflect the new information learned. This operation is performed using a complete set of conditional probabilities $P(\omega|\beta)$---that is, for all $\beta\in\mathcal{B}$ and all $\omega\in\mathcal{O}$---which may be calculated using the ruleset (classical or quantum), the input state, and the interferometer schematic (see \Cref{gamemaster}).
These allow Bayes' theorem to be employed in the form 
\begin{equation}\label{bayes_thm}
P(\beta|\omega)=\frac{P(\beta)P(\omega|\beta)}{\sum_{\beta'}P(\beta')P(\omega|\beta')},
\end{equation}
so that on measuring outcome $\omega_\text{out}$, priors may be updated as $P(\beta)\to P(\beta|\omega_\text{out})$.

As there are only $d+1$ measurement outcomes (four for Hofmann's interferometer), it may seem at first glance feasible to construct an instruction list of optimal moves (photon states) for a player to follow for a given interferometer. 
However, as the outcomes are inherently probabilistic, this list of moves would necessarily branch after each turn.
For the first $n'$ turns, such a list would require the calculation and storage of $\sum_{n=1}^{n'}(d+1)^{n-1}$ initial photons states. For Hofmann's interferometer, where $d=3$ and which typically takes approximately ten turns to complete using some of the better strategies we give below, a table of the first ten optimized moves would contain 349,525 entries, so this is not something we shall pursue in this work.

\subsection{Instructions for the Gamemaster (GM)}
In the physical scenario this game maps to, the player would have to compute the updated priors themself based on detector clicks. However, to help define this scenario as a game, remove trivial losing strategies, and allow us to have a set winning condition, we introduce the Gamemaster (GM), playing the role of the laws of physics. The GM also calculates the updated priors and gives these to the player. The role of the Gamemaster is threefold:
\begin{enumerate}[noitemsep]
\item They play the role of nature, using their knowledge of the blockage location $\beta_\text{secret}$ and a random number generator to decide which detector clicks (or whether none do) according to probability density $P(\omega|\beta_\text{secret})$.
\item They assist the player by performing the above calculations to update their prior probabilities each turn.
\item They use their knowledge of the priors to inform the player when the game is finished.
\end{enumerate}
The protocol the GM follows is summarised in \Cref{gamemaster}.
\begin{algorithm}[H]
  \caption{Protocol for Gamemaster (GM)}
  \label{gamemaster}
\parbox{0.90\linewidth}{
   \begin{algorithmic}[1]
\State initialise secret blockage location, $\beta_\text{secret}$
\State set $\Turn \leftarrow 0$
\State for all $\beta$ set $\Priors(\beta)\leftarrow 1/|\mathcal{B}|$
\While{max($\Priors$) $<\, 1$}
\State set $\Turn \leftarrow \Turn+1$
\State {\bf ask player} to select $\InputState$ with which to take their turn
\State use $\InputState$ to calculate $\ProbDgB(\omega,\beta_\text{secret})$ for all $\omega$ 
\Comment{by employing \Cref{quantum_forward_probs} or \Cref{classical_forward_probs}\hspace{5ex}}
\State set $\omega_\text{out}$ randomly according to $\ProbDgB(\omega,\beta_\text{secret})$
\State use $\InputState$ to calculate $\ProbDgB(\omega_\text{out},\beta)$ for all $\beta$ 
\Comment{by employing \Cref{quantum_forward_probs} or \Cref{classical_forward_probs}\hspace{5ex}}
\State for all $\beta$ set $\ProbBgD(\beta,\omega_\text{out})$
\vspace{-0.8\abovedisplayskip}
\begin{align*}
\leftarrow \frac{\Priors(\beta)\ProbDgB(\omega_\text{out},\beta)}{\sum_{\beta'}\Priors(\beta')\ProbDgB(\omega_\text{out},\beta')}
\end{align*}
\State for all $\beta$ set $\Priors(\beta)\leftarrow \ProbBgD(\beta,\omega_\text{out})$
\State {\bf inform player} of the outcome $\omega_\text{out}$ and updated priors $\Priors$
\EndWhile
\State set $\PlayerGuess\leftarrow \text{arg max}_\beta\left(\Priors(\beta)\right)$
\State {\bf inform player} that the game has concluded, taking $\Turn$ turns to complete and that they have determined the blockage location to be $\PlayerGuess$
   \end{algorithmic}
}
\end{algorithm}

\subsection{Rules for calculating (quantum) probabilities}
\label{quantum_probability_rules}
On each turn and for each possible input photon state considered, the GM protocol (as well as the player protocols for different strategies) (\Cref{gamemaster,player_data_gathering}) require a set of conditional probabilities $P(\omega|\beta)$, which are calculated as follows.
In the path-encoded interferometers considered (see \Cref{sec:interferometer,fig:interferometer,fig:Hofmann_interferometer}), an input photon with density matrix $\rho_\text{Qu}$ undergoes successive two-dimensional unitary transformations performed at each beam splitter, $U_1,U_2,...,U_Q$, where $Q$ refers to the number of beam splitters.
The probability that, if measured, the photon would be found to pass along the $p^\text{th}$ path of the $q^\text{th}$ context may conveniently be expressed as
\begin{equation}
P(\eta^q_p)=\Tr\left[\rho_\text{Qu}\ketbra{\eta^q_p}{\eta^q_p}\right],
\end{equation}
where $\ket{\eta^1_p}=\ket{\alpha_p}$, and for $q\geq2$
\begin{equation}
\ket{\eta^q_p}:=U_1^\dagger\dots U_{q-1}^\dagger\ket{\alpha_p}.
\end{equation}
The first context $\{\ket{\eta^1_1},...,\ket{\eta^1_d}\}$ corresponds to the input states prior to the photons arriving at the first beam splitter, and the detector states comprise the $(Q+1)^\text{th}$ context $\left\{\ket{\eta^{Q+1}_1},...,\ket{\eta^{Q+1}_d}\right\}$.
As all possible blockage locations exist on a segment of path joining two adjacent nodes, for all such locations $\beta$ there exists a corresponding state $\ket{\beta}$ equal to $\ket{\eta^q_p}$ for at least one pair $(q,p)$.
In Hofmann's interferometer (see \Cref{fig:Hofmann_interferometer}) the blockage locations $\beta$ are labelled $D_1,S_1,P_1,F,P_2,S_2,$ and $D_2$, where for instance the state $\ket{D_1}$ is equal to $\ket{\eta_2^2}$, and $\ket{S_2}$ is equal to $\ket{\eta^4_3}$ or $\ket{\eta^5_3}$.
To avoid repeating calculations, the components of these blockage vectors may be tabulated prior to the game.

The photon detectors $\omega_p$ ($p\geq 1$) are associated with states $\ket{\omega_p}=\ket{\eta^{Q+1}_p}$, where $Q$ is the number of beam splitters in the interferometer, so that in the absence of a blockage the probability that during a turn the $p^\text{th}$ detector activates is
\begin{equation}
P(\omega_p|\text{no blockage})=\Tr\left[\rho_\text{Qu}\ketbra{\omega_p}{\omega_p}\right].
\end{equation}
The presence of a blockage $\beta\in\mathcal{B}$ modifies this density to 
\begin{widetext}
\begin{equation}
P(\omega_p|\beta)=
\begin{cases}
\tr\Big[\left(\mathds{1}-\ketbra{\beta}{\beta}\right)\rho_\text{Qu}\left(\mathds{1}-\ketbra{\beta}{\beta}\right)\ketbra{\omega_p}{\omega_p}\Big] & p=1,...,d\\
1-\sum_{p=1}^{d}P(\omega_p|\beta) & p=0
\end{cases}.\label{quantum_forward_probs}
\end{equation}
\end{widetext}
For a pure state $\rho_\text{Qu}=\ketbra{\psi}{\psi}$ this simplifies to 
\begin{equation}
P(\omega_p|\beta)=
\begin{cases}
\left|\bra{\omega_p}\left(\mathds{1}-\ketbra{\beta}{\beta}\right)\ket{\psi}\right|^2 & p=1,...,d\\
\left|\braket{\beta}{\psi}\right|^2 & p=0
\end{cases}.\label{quantum_forward_probs_pure_state}
\end{equation}
Inspecting \Cref{quantum_forward_probs,quantum_forward_probs_pure_state}, we note that if there are two blockage locations, $\beta_1$ and $\beta_2$, with the same state vector up to a complex phase, $\ket{\beta_1}=e^{i\theta}\ket{\beta_2}$, then these locations yield identical conditional probabilities and are thus indistinguishable to the player under the quantum ruleset. (Note this is not the case for the classical ruleset.) As it is impossible to complete a game under such conditions, we shall assume $|\braket{\beta}{\beta'}|^2<1$ for all $\beta\neq\beta'$.

\subsection{Alternative ruleset for classical probabilities}

\begin{figure*}
\input{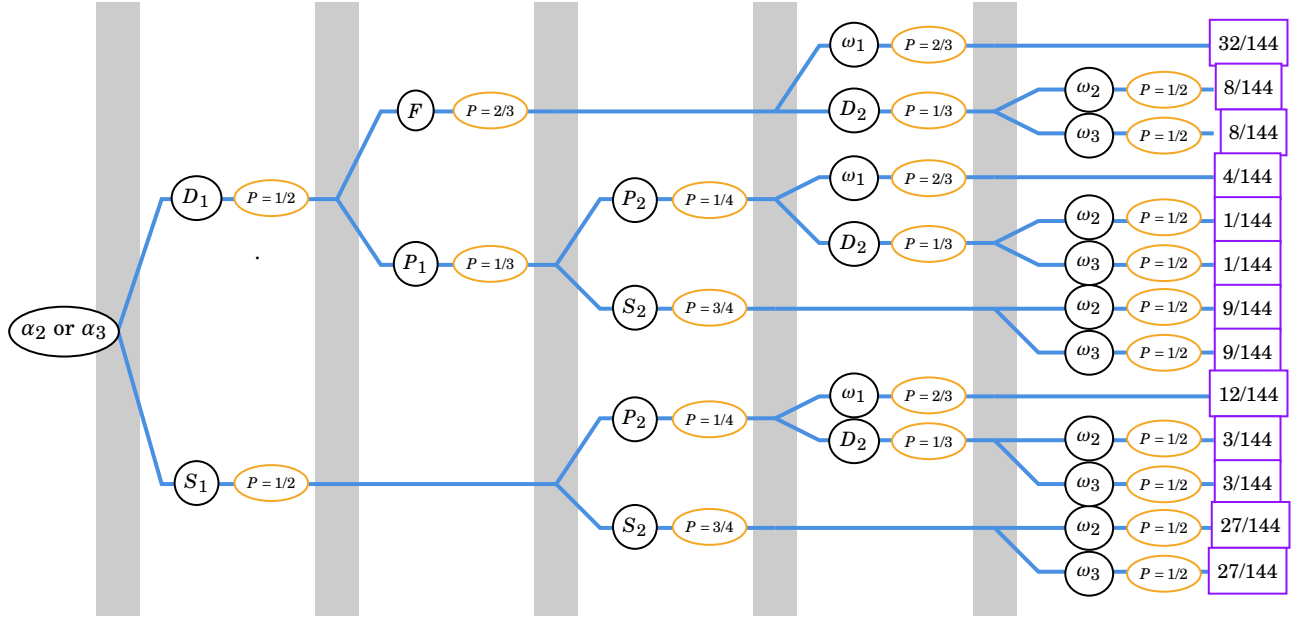}
\caption{Example probability tree for classical values of $P(\omega|\text{no blockage})$ given a photon entered into input $\alpha_2$ or $\alpha_3$. Trees for $\alpha_2$ and $\alpha_3$ are the same due to the $1/2$ reflectivity of the first beamsplitter, $R_1$. In the presence of a blockage at any of the locations $\beta\in\{D_1,S_1,P_1,F,P_2,S_2,D_2\}$, the corresponding branches may be cut off, with the sum over total probabilities at the cuts comprising $P(\omega_0|\beta)$. }
\label{tree_diagram}
\end{figure*}

We may quantify the advantage that quantum information gain has over classical information gain with an alternative classical ``pinball'' ruleset, in which:
\begin{enumerate}
\item Photons are considered to have definite locations in the interferometer at all times. (The player may nevertheless be uncertain of this location.)
\item The player may specify an input probability density (the classical analogue of the density matrix, \Cref{eq:density_matrix}), 
\begin{equation}\label{eq:probability_vector}
\rho_\text{Cl}=
\begin{pmatrix}
\rho_1\\
\vdots\\
\rho_d
\end{pmatrix}
,\quad
\rho_p\in\mathds{R},\quad
\rho_p\geq0,\quad
\sum_{p=1}^d\rho_p=1,
\end{equation}
or the equivalent of a pure state by selecting an input port $p'$ to enter the photon into, equivalent to $\rho_p=\delta_{pp'}$ in \Cref{eq:probability_vector}. 
As with the Schr\"{o}dinger picture quantum analogue, the blockage will act to reduce the normalisation of vector $\rho_\text{Cl}$.
\item Photons reflect probabilistically off beam splitters with the same reflection coefficients as their quantum analogues.
\item A photon is absorbed if it travels along the path segment containing the blockage.
\item If they reach the $p^\text{th}$ output path, the corresponding detector activates, giving output $\omega_p$.
\end{enumerate}

Conditional probabilities $P(\omega|\beta)$ may be calculated with a traditional tree diagram (see \Cref{tree_diagram}), however the branching possibilities make these unwieldy for larger interferometers.
For manageable calculations, instead consider that the effect of a reflector between adjacent paths, with reflectivity $R$, on a classical probability vector $\rho_\text{Cl}$, is to transform it as
\begin{equation}
\rho_\text{Cl}\longrightarrow T\rho_\text{Cl}=
\begin{pmatrix}
\ddots&&\\
&R&1-R&\\
&1-R&R&\\
&&&\ddots
\end{pmatrix}
\rho_\text{Cl}.
\end{equation}
In Hofmann's interferometer (\Cref{fig:Hofmann_interferometer}), the beam splitters are labelled in time order as $1,S_1,F,S_2$, and $2$, with respective reflection coefficients $R_1=1/2$, $R_{S_1}=1/3$, $R_F=1/4$, $R_{S_2}=1/3$, and $R_2=1/2$. Denoting the corresponding transition matrices $T_1,T_{S_1},T_F,T_{S_2},T_{2}$, so that for instance
\begin{equation}
T_{S_1}=
\begin{pmatrix}
1/3&2/3&0\\
2/3&1/3&0\\
0&0&1
\end{pmatrix}
,\quad T_F=
\begin{pmatrix}
1&0&0\\
0&1/4&3/4\\
0&3/4&1/4
\end{pmatrix},
\end{equation}
the probability of detector $p$ activating upon a given input photon density $\rho_\text{Cl}$ without a blockage is
\begin{equation}
P\left(\omega_p|\text{no blockage}\right)
=\left(T_2T_{S_2}T_FT_{S_1}T_1\rho_\text{Cl}\right)_p.
\end{equation}

A blockage on the $k^\text{th}$ path transforms the probability density as $\rho_\text{Cl}\to X_k\rho_\text{Cl}$, where matrix $X_k$ has components $(X_k)_{ij}=\delta_{ij}-\delta_{ik}\delta_{jk}$.
By inspection of \Cref{fig:Hofmann_interferometer}, we find that in the presence of each blockage the probability of the $p^\text{th}$ detector activating may be expressed
\begin{equation}
\begin{split}
P\left(\omega_p|\beta=D_1\right)=\left(T_2T_{S_2}T_FT_{S_1}X_2T_1\rho_\text{Cl}\right)_p\\
P\left(\omega_p|\beta=S_1\right)=\left(T_2T_{S_2}T_FT_{S_1}X_3T_1\rho_\text{Cl}\right)_p\\
P\left(\omega_p|\beta=P_1\right)=\left(T_2T_{S_2}T_FX_2T_{S_1}T_1\rho_\text{Cl}\right)_p\\
P\left(\omega_p|\beta=F\right)=\left(T_2T_{S_2}T_FX_1T_{S_1}T_1\rho_\text{Cl}\right)_p\\
P\left(\omega_p|\beta=P_2\right)=\left(T_2T_{S_2}X_2T_FT_{S_1}T_1\rho_\text{Cl}\right)_p\\
P\left(\omega_p|\beta=S_2\right)=\left(T_2T_{S_2} X_3T_FT_{S_1}T_1\rho_\text{Cl}\right)_p\\
P\left(\omega_p|\beta=D_2\right)=\left(T_2X_2T_{S_2}T_FT_{S_1}T_1\rho_\text{Cl}\right)_p
\end{split},\label{classical_forward_probs}
\end{equation}
where as before $P(\omega_0|\beta)=1-\sum_{p=1}^dP(\omega_p|\beta)$.
The result of these calculations for definite (classically pure) states entered into input port $\alpha_1,\alpha_2$, or $\alpha_3$ are listed in \Cref{table:classical_probs}.
Note that, as the first reflector ($D_1$) has reflectivity $R_1=1/2$, the tables of $P(\omega|\beta)$ are identical for photons entering through either ports $\alpha_2$ or $\alpha_3$.
Outcome conditional probabilities for input photons with uncertain initial states $\rho_\text{Cl}=(\rho_1,\rho_2,\rho_3)^T$ may be accounted for with a weighted average over the values in \Cref{table:classical_probs}.

\begin{table}
\begin{tabular}[c]{|c|c|c|c|c|}
\hline 
Blockage position $\displaystyle \beta $
 & 
$\displaystyle P( \omega _{1} |\beta )$
 &
$\displaystyle P( \omega _{2} |\beta )$
 &
$\displaystyle P( \omega _{3} |\beta )$
 &
$\displaystyle P( \omega _{0} |\beta )$
 \\
\hline 
$\displaystyle D_{1}$
 &
$\displaystyle 0.222$
 &
$\displaystyle 0.389$
 &
$\displaystyle 0.389$
 &
$\displaystyle 0$
 \\
\hline 
$\displaystyle S_{1}$
 &
$\displaystyle 0.222$
 &
$\displaystyle 0.389$
 &
$\displaystyle 0.389$
 &
$\displaystyle 0$
 \\
\hline 
$\displaystyle P_{1}$
 &
$\displaystyle 0.111$
 &
$\displaystyle 0.111$
 &
$\displaystyle 0.111$
 &
$\displaystyle 0.667$
 \\
\hline 
$\displaystyle F$
 &
$\displaystyle 0.111$
 &
$\displaystyle 0.278$
 &
$\displaystyle 0.278$
 &
$\displaystyle 0.333$
 \\
\hline 
$\displaystyle P_{2}$
 &
$\displaystyle 0.111$
 &
$\displaystyle 0.361$
 &
$\displaystyle 0.361$
 &
$\displaystyle 0.167$
 \\
\hline 
$\displaystyle S_{2}$
 &
$\displaystyle 0.222$
 &
$\displaystyle 0.139$
 &
$\displaystyle 0.139$
 &
$\displaystyle 0.500$
 \\
\hline 
$\displaystyle D_{2}$
 &
$\displaystyle 0.222$
 &
$\displaystyle 0.250$
 &
$\displaystyle 0.250$
 &
$\displaystyle 0.278$
 \\
\hline 
No blockage
 &
$\displaystyle 0.222$
 &
$\displaystyle 0.389$
 &
$\displaystyle 0.389$
 &
$\displaystyle 0$
 \\
 \hline
\end{tabular}\\
\vspace{5pt}
\begin{tabular}[c]{|c|c|c|c|c|}
\hline 
Blockage position $\displaystyle \beta $
 & 
$\displaystyle P( \omega _{1} |\beta )$
 &
$\displaystyle P( \omega _{2} |\beta )$
 &
$\displaystyle P( \omega _{3} |\beta )$
 &
$\displaystyle P( \omega _{0} |\beta )$
 \\
\hline 
$\displaystyle D_{1}$
 &
$\displaystyle 0.250$
 &
$\displaystyle 0.125$
 &
$\displaystyle 0.125$
 &
$\displaystyle 0.500$
 \\
\hline 
$\displaystyle S_{1}$
 &
$\displaystyle 0.139$
 &
$\displaystyle 0.181$
 &
$\displaystyle 0.181$
 &
$\displaystyle 0.500$
 \\
\hline 
$\displaystyle P_{1}$
 &
$\displaystyle 0.361$
 &
$\displaystyle 0.236$
 &
$\displaystyle 0.236$
 &
$\displaystyle 0.167$
 \\
\hline 
$\displaystyle F$
 &
$\displaystyle 0.278$
 &
$\displaystyle 0.194$
 &
$\displaystyle 0.194$
 &
$\displaystyle 0.333$
 \\
\hline 
$\displaystyle P_{2}$
 &
$\displaystyle 0.111$
 &
$\displaystyle 0.236$
 &
$\displaystyle 0.236$
 &
$\displaystyle 0.417$
 \\
\hline 
$\displaystyle S_{2}$
 &
$\displaystyle 0.389$
 &
$\displaystyle 0.181$
 &
$\displaystyle 0.181$
 &
$\displaystyle 0.250$
 \\
\hline 
$\displaystyle D_{2}$
 &
$\displaystyle 0.389$
 &
$\displaystyle 0.125$
 &
$\displaystyle 0.125$
 &
$\displaystyle 0.0.2501$
 \\
 \hline
 No blockage
 &
$\displaystyle 0.389$
 &
$\displaystyle 0.306$
 &
$\displaystyle 0.306$
 &
$\displaystyle 0$
 \\
 \hline
\end{tabular}
\caption{Above: Probabilities of each outcome (out of 144) in Hofmann's interferometer given a ``classical photon'' has been entered into input port $\alpha_1$.
Below: The same for classical photons entered into either input port $\alpha_2$ or $\alpha_3$.}
\label{table:classical_probs}
\end{table}

\section{Analysis of gameplay and strategy}\label{sec:GameplayAnalysis}
\subsection{Metrics for game progress}
For non-trivial interferometer setups, the number of blockage locations $|\mathcal{B}|$ is generally substantially larger than the number of possible outcomes $|\mathcal{O}|=d+1$, meaning it is unlikely that it will be possible to finish a game in one turn.
For this reason the player benefits from having a metric with which they can measure game progress, and around which they can build strategies. We have primarily based our strategies around two metrics, which we now explain. The first of these is countable and increases monotonically, the second is continuous and has the potential to decrease.

Our first way to measure game progress is to track the number of blockages ruled out.
That is, rather than attempting determine where the blockage \emph{is}, the player may attempt to determine where it \emph{isn't}.
\Cref{ruling_out_blockages} describes various methods with which to rule out blockages.
However, for the quantum ruleset a simple strategy that aims to rule out blockage locations one-by-one may be formulated with contrapositional logic. If for a given blockage $\beta$, the player chooses $\ket{\psi}=\ket{\beta}$, then by \Cref{quantum_forward_probs} $P(\omega_0|\beta)=1$, so that if the blockage is at location $\beta$, the photon is guaranteed to be blocked. This does not mean that the outcome $\omega_0$ confirms the location $\beta$, as the photon may have been blocked by a different blockage (which can occur when $P(\beta|\omega_0)$ doesn't equal $P(\omega_0|\beta))$. However, as $P(\omega_0|\beta)=1$ is equivalent to the implication $\beta\implies\omega_0$, which by contraposition becomes $\neg\omega_0\implies\neg\beta$, if the photon is not absorbed then the player may rule out location $\beta$.
By following this logic, a player may ignore all Bayesian inference, and need not perform any calculations, but instead simply needs to keep a list of the blockages yet to be ruled out: 
\begin{strategy}\label{simple_strategy}
On each turn, randomly choose a blockage location $\beta$ among the remaining locations that are yet to be ruled out, i.e. for which $P(\beta)>0$. Set the input state equal to the blockage location state, $\ket{\psi}=\ket{\beta}$. If a detector activates, rule out this blockage location. 
Repeat until only one blockage location remains.
\end{strategy}
Ruling out a location in this way is an instance of conclusive state exclusion~\mbox{\cite{Caves2002Compatibility,Bandyopadhyay2014Exclusion}}: the outcome certifies that the photon was not blocked at $\beta$, without certifying where it was. Ruling out several locations in a single turn, as discussed in \mbox{\Cref{ruling_out_blockages}}, corresponds to the antidistinguishability of the corresponding set of blockage states~\mbox{\cite{Heinosaari2018Antidistinguishability}}.

\begin{figure}
\input{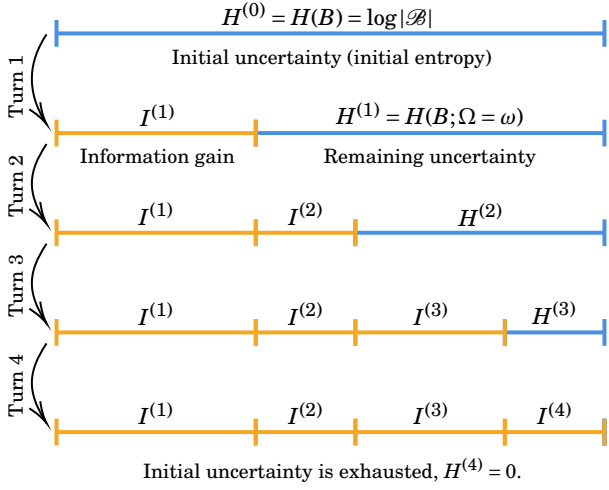}
\caption{Information gain in an example game that takes 4 turns. The initial uncertainty $\log|\mathcal{B}|$ is exhausted by the information gained over the 4 turns. }
\label{recursive_information_gain}
\end{figure}

A more granular method to measure progress is in terms of the (information) entropy of the blockage probabilities $P(\beta)$, or as we explain below the entropic ``equivalent number'' of blockages ruled out. 
At the beginning of the game, the player's priors, $P(\beta)=1/|\mathcal{B}|$ for all $\beta$, reflect the principle of indifference (see for instance Ref.~\cite{jaynes_probability_2003}) and correspond to maximum entropy for the situation, 
\begin{equation}
H(B)=-\sum_{\beta\in\mathcal{B}}P(\beta)\log P(\beta)=\log|\mathcal{B}|.
\end{equation}
The game ends when the priors are $P(\beta)=\delta(\beta,\gamma)$ for some $\gamma\in\mathcal{B}$ (returning unity for $\beta=\gamma$, and zero otherwise), which corresponds to minimum entropy, $H(B)=0$.
Upon taking the first turn and recording the detector outcome $\omega_\text{out}$, the remaining uncertainty in the blockage location is quantified by the posterior entropy,
\begin{equation}
H(B|\Omega\mathord{=}\omega_\text{out})=-\sum_{\beta\in\mathcal{B}}P(\beta|\omega_\text{out})\log P(\beta|\omega_\text{out}).
\end{equation}
We can measure the effectiveness of the turn with the \emph{Information Gain}\footnote{We use terminology common in the field of Bayesian Experimental Design \cite{rainforthModernBayesianExperimental2024,ryanReviewModernComputational2016,sebastianiMaximumEntropySampling2000}. Quantities $H(B|\Omega=\omega_\text{out})$ and $I(B;\Omega=\omega_\text{out})$ are referred to as the (actual) posterior entropy and information gain, as distinguished these from their expectation value analogues, $H(B|\Omega)$ and $I(B;\Omega)$. Note that it is common to refer to the expected posterior entropy as the conditional entropy, and to refer to the expected information gain as the mutual information.} $I(B;\Omega\mathord{=}\omega)$, defined such that 
\begin{equation}
\begin{split}
H(B)&=I(B;\Omega\mathord{=}\omega_\text{out})+H(B|\Omega\mathord{=}\omega_\text{out}),
\end{split}
\end{equation}
i.e.,
\begin{equation}
\begin{split}
\begin{matrix}
\text{initial}\\
\text{uncertainty}
\end{matrix}
\,\,\,&=\,\,\,
\begin{matrix}
\text{information}\\
\text{gained}
\end{matrix}
\,\,\,+\,\,\,
\begin{matrix}
\text{remaining}\\
\text{uncertainty}
\end{matrix}\,\,\,.
\end{split}
\end{equation}
When this first trial is complete, and the prior probabilities are updated, $P(\beta)\to P(\beta|\omega_\text{out})$, the process may be repeated with a new input photon state, and the player may use entropy to track overall game progress. If in each new trial information is gained, then it is possible that a trial will eventually be reached when the cumulative information gain exhausts the uncertainty (entropy), at which point the remaining uncertainty will be zero, and the game will be complete. Denoting the information gain of the $n^\text{th}$ trial $I^{(n)}$, this will happen if for some $N\in\mathbb{N}$,
\begin{equation}
\log|\mathcal{B}| = I^{(1)}+I^{(2)}+...+I^{(N)}.
\end{equation}
This recursive exhaustion of the initial uncertainty is depicted in \Cref{recursive_information_gain}. Of course there is no guarantee that such an $N^\text{th}$ trial will ever be reached--see the below discussion on why an information gain optimizer algorithm described in \Cref{info_gain_optimization} fails to complete the game when using a brute force sampling method.
Additionally, note that it is possible (and potentially necessary in some circumstances) to ``lose information'', meaning obtain a negative information gain. 
We address this phenomenon further in the discussion.

To better understand what information gain means in terms of the blockages, note that there will be some real number $PP\in[1,|\mathcal{B}|]$ (which is unlikely to be an integer on any mid-game turn) such that
\begin{equation}
H(B|\Omega\mathord{=}\omega_\text{out})=\log PP,
\end{equation}
so that we may (somewhat loosely) regard the remaining uncertainty post-trial to be informationally equivalent to complete uncertainty over $PP$ blockages.\footnote{We borrow the notation $PP$ from Ref.~\cite{jelinek_perplexitymeasure_2005} who name it the ``perplexity''.} Loosely, this means the information gained in the first turn is informationally equivalent to ruling out $|\mathcal{B}|-PP$ blockage locations.
Denoting the entropy prior to the $n^\text{th}$ trial $H^{(n-1)}$, and taking the entropy logarithm base equal to 2, the number of additional blockages ruled out in the $n^\text{th}$ trial is
\begin{equation}
2^{H^{(n-1)}}-2^{H^{(n)}}.
\end{equation}
The game is completed when the sum of these totals $|\mathcal{B}|-1$, indicating there is only one blockage remaining,
\begin{equation}
\sum_{n=1}^m\left(2^{H^{(n-1)}}-2^{H^{(n)}}\right)
=2^{H^{(0)}}-2^{H^{(N)}}
=|\mathcal{B}|-1.
\end{equation}
\Cref{info_gain_optimization} describes a strategy that aims to maximise information gain.

\subsection{Quantum Ruleset: How to rule out blockages}\label{ruling_out_blockages}
We begin our move analysis by considering only pure input states, $\rho_\text{Qu}=\ketbra{\psi}{\psi}$, under the quantum ruleset.
Under this restriction we can use \Cref{bayes_thm,quantum_forward_probs_pure_state} to give the conditional probability of blockage $\beta$, given that input photon $\ket{\psi}$ is blocked (outcome $\omega_0$), as
\begin{equation}\label{backward_quantum_probs_absorbed}
P(\beta|\omega_0)=\frac{|\braket{\beta}{\psi}|^2P(\beta)}{\sum_{\beta'}|\braket{\beta'}{\psi}|^2P(\beta')}.
\end{equation}
Here we see that, to be able to obtain $P(\beta|\omega_0)=1$ for some $\beta$, and so finish the game on a turn in which the photon is blocked, the player must have chosen an input state $\ket{\psi}$ that is orthogonal to all blockage vectors for blockages yet to be ruled out, except for that of the correct location. 
This requires the player not only to have correctly guessed the blockage location, but also that the other locations span an orthogonal (hyper)plane.
While it is reasonable to expect some blockage vectors to be orthogonal---for instance blockage vectors belonging to the same context are necessarily orthogonal---this is rarely feasible except for in the late game.
As such, it is typically more useful to attempt to advance game progress by aiming to rule out blockages, through finding a $\ket{\psi}$ for which $P(\beta|\omega_0)=0$ for some $\beta$.

From \Cref{backward_quantum_probs_absorbed}, we see that when the photon is absorbed, a blockage $\beta$ may be ruled out when $\braket{\beta}{\psi}=0$ (which, by Hilbert space symmetry, implies $\braket{\psi}{\beta}=0$).
For a single $\beta$, choosing a $\ket{\psi}$ such that $\braket{\psi}{\beta}=0$ is not difficult.
A simple way to achieve this is to select $\ket{\psi}=\ket{\beta'}$ for a blockage location $\beta'$ within the same context as $\beta$.
If such a photon state is blocked then all other blockages within the same context ($\neq\beta'$) may be ruled out (if this hasn't already been done), and if the photon is not blocked, the contrapositional logic of Strategy~\ref{simple_strategy} may be used as a fallback to rule out blockage $\beta'$ (again, if this hasn't already happened).
This property of being naturally orthogonal to $d-1$ other locations, and parallel to one, makes blockage vectors a good choice of input state for a strategy based on ruling out blockage locations.

A player wishing to follow a strategy of ruling out blockages may, prior to playing the game, construct tables of $P(\omega|\beta)$.
Unlike the conditionals $P(\beta|\omega)$, these are independent of priors $P(\beta)$, and so valid for the whole game. However, it is conveniently the case that $P(\omega|\beta)=0\implies P(\beta|\omega)=0$, meaning identifying cases where $P(\omega|\beta)=0$ for an initial state allows us to identify guaranteed cases where $P(\beta|\omega)=0$ for that input state, regardless of the priors $P(\beta)$.
We list such tables of $P(\omega|\beta)$s for Hofmann's interferometer in Appendix~\ref{info_tables}, for $\ket{\psi}=\ket{\beta}$ for all $\beta$, for $\ket{\psi}=\ket{\omega_1},...,\ket{\omega_3}$\footnote{Note Hofmann's interferometer has the useful property that the output context $\{\ket{\omega_i}\}_i$ is the same as the input context $\{\ket{\alpha_i}\}_i$ by design.}, and for certain other input states (informed by the analysis in Ref.~\mbox{\cite{Hofmann2025Sphere}}).
To help optimize a strategy based on ruling out blockages in the case of photon absorption, the player may want to consider the zeros in the top row of each table.
Disregarding for a moment the state of the priors, we see that input photon states equal to $\ket{S_1}$ and $\ket{S_2}$ are both generally good choices, able to rule out three locations each. 
However, choosing $\ket{\psi}=\ket{F}$ appears optimal, able to rule out 4 blockages ($S_1,P_1,P_2$, and $S_2$) on outcome $\omega_0$. We shall elaborate on this in \Cref{why_F_is_optimal}.

An ambitious player may want to try ruling out multiple specific blockages in a single turn, or to rule out as many locations as possible. 
For the former, given a set of locations $\beta_1,...,\beta_l$, the player should first verify that $\ket{\beta_1},...,\ket{\beta_l}$ span at most a $k\leq d-1$ dimensional subspace of the Hilbert space. 
This could be done by verifying that a matrix formed from the columns of vectors $\ket{\beta_1},...,\ket{\beta_l}$ has a rank of $k\leq d-1$, for instance by checking that the size of the largest square sub-matrix with non-zero determinant is $k\times k$.
For the latter, the player may look for the largest subset $\{\beta_1,...,\beta_l\}\subset\mathcal{B}$ such that $\ket{\beta_1},...,\ket{\beta_l}$ spans at most a $k\leq d-1$ dimensional subspace. 
In either case, the player may proceed by first finding a subset of $k\leq l$ of these vectors, where all $k$ vectors are linearly independent, verifying this by checking that the matrix formed from the columns of these $k$ vectors is of rank $k$.
Then, by taking these vectors as equal to $\ket{v_1},...,\ket{v_k}$ in the following procedure, the player can obtain an input state $\ket{\psi'}$ orthogonal to all $\ket{\beta_1},...,\ket{\beta_l}$, so that $P(\omega_0|\beta_i)=0$ for all $i=1,..,l$.
If such an input state $\ket{\psi'}$ is used, and the photon is absorbed, then all of these blockage locations $\beta_1,...,\beta_l$ may be ruled out. 

\begin{procedure}\label{gram_schmidt_procedure}
Let $\ket{v_1},...,\ket{v_k}$ be $k\leq d-1$ linearly independent vectors, and let $\ket{\psi}$ be a vector that is linearly independent to these.
Define 
\begin{equation}\label{GS}
\begin{split}
\ket{v'_1}&:=\ket{v_1},\\
\ket{v'_2}&:=\left(\mathds{1}-\frac{
\ketbra{v'_1}{v'_1}
}{
\braket{v'_1}{v'_1}
}\right)\ket{v_2},\\
&\,\,\,\,\,\,\vdots\\
\ket{v'_k}&:=\left(\mathds{1}-\sum_{i=1}^{k-1}\frac{
\ketbra{v'_i}{v'_i}
}{
\braket{v'_i}{v'_i}
}\right)\ket{v_k}.
\end{split}
\end{equation}
This is a typical Gram-Schmidt set of orthogonal vectors. By construction, $\braket{v'_i}{v'_j}=0$ for all $i\neq j$.
Now define 
\begin{equation}
\ket{\psi'}:=N\left(\mathds{1}-\sum_{i=1}^{k}\frac{
\ketbra{v'_i}{v'_i}
}{
\braket{v'_i}{v'_i}
}\right)\ket{\psi},
\end{equation}
where $N$ is a normalization factor introduced such that $\braket{\psi'}{\psi'}=1$.
This satisfies 
\begin{equation}
\begin{split}
\braket{v'_j}{\psi'}
&=N\bra{v'_j}
\left(\mathds{1}-\sum_{i=1}^{k}\frac{
\ketbra{v'_i}{v'_i}
}{
\braket{v'_i}{v'_i}
}\right)\ket{\psi}\\
&=N\braket{v'_j}{\psi}
-
N\frac{
\braket{v'_j}{v'_j}
}{
\braket{v'_j}{v'_j}}\braket{v'_j}{\psi}=0
\end{split}
\end{equation}
for all $j$.
Now, from \Cref{GS} we can see that we may also expand out all $\bra{v_j}$ in terms of the Gram-Schmidt vectors. That is, for any $1\leq j\leq k$,
\begin{equation}
\bra{v_j}=\bra{v'_j}+\sum_{i=1}^{j-1}
\frac{
\braket{v_j}{v'_i}
}{
\braket{v'_i}{v'_i}}\bra{v'_i}.
\end{equation}
Thus $\braket{v_j}{\psi'}=0$ for all $1\leq j\leq k$.
\end{procedure}

So far in this section, we have only considered the ruling out of blockages when the photon is absorbed, outcome $\omega_0$.
A similar analysis may be performed for detector activations, outcomes $\omega_p$ for $p=1,...,d$.
Given detector $p$ activates, the probability $P(\beta|\omega_p)$ that the blockage location is $\beta$ is
\begin{equation}\label{backward_quantum_probs_detected}
P(\beta|\omega_p)=
\frac{|\bra{\omega_p}\left(\mathds{1}-\ketbra{\beta}{\beta}\right)\ket{\psi}|^2P(\beta)}{\sum_{\beta'}|\bra{\omega_p}\left(\mathds{1}-\ketbra{\beta'}{\beta'}\right)\ket{\psi}|^2P(\beta')}.
\end{equation}
A first point to note is that if $\ket{\psi}$ is chosen to be equal to $\ket{\beta}$ for some $\beta$, then $P(\beta|\omega_p)$ vanishes for all $p=1,...,d$, so no detector can activate. Thus if any detector does activate, location $\beta$ may be ruled out. 
This is an alternative argument for Strategy~\ref{simple_strategy} above.
If detector $\omega_p$ activates, then one may scan the corresponding row in the $\ket{\psi}=\ket{\beta}$ table of $P(\omega|\beta)$ for zeros to determine which blockages may be ruled out. (See Appendix~\ref{info_tables} for these tables for Hofmann's interferometer.)

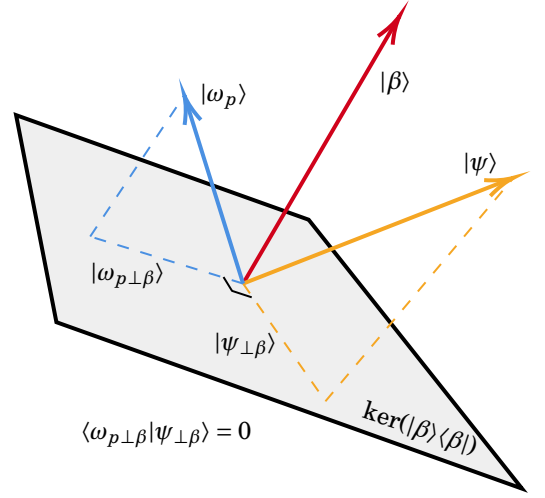
\begin{figure}
    \centering
\tikzset{every picture/.style={line width=0.75pt}} 

\begin{tikzpicture}[x=0.75pt,y=0.75pt,yscale=-1,xscale=1]

\draw  [color={rgb, 255:red, 0; green, 0; blue, 0 }  ,draw opacity=1 ][fill={rgb, 255:red, 239; green, 239; blue, 239 }  ,fill opacity=1 ][line width=1.5]  (60.37,158.63) -- (40.13,54.76) -- (186.97,107.23) -- (294.05,242.12) -- cycle ;
\draw [color={rgb, 255:red, 208; green, 2; blue, 27 }  ,draw opacity=1 ][line width=1.5]    (154.07,139.27) -- (231.11,7.59) ;
\draw [shift={(232.62,5)}, rotate = 120.33] [color={rgb, 255:red, 208; green, 2; blue, 27 }  ,draw opacity=1 ][line width=1.5]    (14.21,-4.28) .. controls (9.04,-1.82) and (4.3,-0.39) .. (0,0) .. controls (4.3,0.39) and (9.04,1.82) .. (14.21,4.28)   ;
\draw [color={rgb, 255:red, 245; green, 166; blue, 35 }  ,draw opacity=1 ][line width=1.5]    (154.07,139.27) -- (287.04,87.24) ;
\draw [shift={(289.83,86.15)}, rotate = 158.63] [color={rgb, 255:red, 245; green, 166; blue, 35 }  ,draw opacity=1 ][line width=1.5]    (14.21,-4.28) .. controls (9.04,-1.82) and (4.3,-0.39) .. (0,0) .. controls (4.3,0.39) and (9.04,1.82) .. (14.21,4.28)   ;
\draw [color={rgb, 255:red, 245; green, 166; blue, 35 }  ,draw opacity=1 ] [dash pattern={on 4.5pt off 4.5pt}]  (195.39,198.3) -- (289.83,86.15) ;
\draw [color={rgb, 255:red, 245; green, 166; blue, 35 }  ,draw opacity=1 ] [dash pattern={on 4.5pt off 4.5pt}]  (154.07,139.27) -- (195.39,198.3) ;
\draw [color={rgb, 255:red, 74; green, 144; blue, 226 }  ,draw opacity=1 ][line width=1.5]    (154.07,139.27) -- (125.45,47.69) ;
\draw [shift={(124.55,44.83)}, rotate = 72.65] [color={rgb, 255:red, 74; green, 144; blue, 226 }  ,draw opacity=1 ][line width=1.5]    (14.21,-4.28) .. controls (9.04,-1.82) and (4.3,-0.39) .. (0,0) .. controls (4.3,0.39) and (9.04,1.82) .. (14.21,4.28)   ;
\draw [color={rgb, 255:red, 74; green, 144; blue, 226 }  ,draw opacity=1 ] [dash pattern={on 4.5pt off 4.5pt}]  (77.33,115.66) -- (124.55,44.83) ;
\draw [color={rgb, 255:red, 74; green, 144; blue, 226 }  ,draw opacity=1 ] [dash pattern={on 4.5pt off 4.5pt}]  (154.07,139.27) -- (77.33,115.66) ;
\draw    (148.36,142.87) -- (158.39,146.16) ;
\draw    (144.23,135.98) -- (148.56,143.01) ;

\draw (221,31.4) node [anchor=north west][inner sep=0.75pt]  [font=\normalsize]  {$| \beta \rangle $};
\draw (215.36,195.8) node [anchor=north west][inner sep=0.75pt]  [font=\normalsize,rotate=-18.22]  {$\text{ker}(| \beta \rangle \langle \beta |)$};
\draw (263,72.4) node [anchor=north west][inner sep=0.75pt]  [font=\normalsize]  {$| \psi \rangle $};
\draw (131,37.4) node [anchor=north west][inner sep=0.75pt]  [font=\normalsize]  {$| \omega _{p} \rangle $};
\draw (76,127.4) node [anchor=north west][inner sep=0.75pt]  [font=\normalsize]  {$| \omega _{p\perp \beta } \rangle $};
\draw (138,162.4) node [anchor=north west][inner sep=0.75pt]  [font=\normalsize]  {$| \psi _{\perp \beta } \rangle $};
\draw (71,207.4) node [anchor=north west][inner sep=0.75pt]  [font=\normalsize]  {$\langle \omega _{p\perp \beta }| \psi _{\perp \beta } \rangle =0$};

\end{tikzpicture}
    \caption{Illustration of why blockage $\beta$ may be ruled out when detector $\omega_p$ (for $p\neq0$) activates, if the inner product between the projection of both $\ket{\psi}$ and $\ket{\omega_p}$ onto the kernel of $\ketbra{\beta}{\beta}$ vanishes.}
    \label{inner_product_of_projections}
\end{figure}

More generally, to rule out blockage locations when detector $p$ activates, we should consider the conditions under which 
\begin{equation}\label{forward_prob_zero_detected}
\bra{\omega_p}\left(\mathds{1}-\ketbra{\beta}{\beta}\right)\ket{\psi}=0.
\end{equation}
The operator $\mathds{1}-\ketbra{\beta}{\beta}$, which we henceforth write $P_{\perp\beta}$, is an orthogonal projection onto the kernel (null space) of $\ketbra{\beta}{\beta}$.
Due to its idempotency and Hermiticity, we may write the condition $P(\beta|\omega_p)=0$ equivalently as 
\begin{equation}\label{condition_for_ruling_out_given_detector_activation}
0=\braket{\omega_{p\perp\beta}}{\psi}
=\braket{\omega_{p}}{\psi_{\perp\beta}}
=\braket{\omega_{p\perp\beta}}{\psi_{\perp_\beta}},
\end{equation}
where we have defined $\ket{\omega_{p\perp\beta}}:=P_{\perp\beta}\ket{\omega_p}$ and $\ket{\psi_{\perp_\beta}}:=P_{\perp\beta}\ket{\psi}$.
Condition \eqref{condition_for_ruling_out_given_detector_activation} is illustrated in \Cref{inner_product_of_projections}.

Given a $\ket{\psi}$ for which Condition \eqref{condition_for_ruling_out_given_detector_activation} does not hold\footnote{We also require $|\braket{\omega_{p\perp\beta}}{\psi}|^2\neq1$, so that $\left(\mathds{1}-\ketbra{\omega_{p\perp\beta}}{\omega_{p\perp\beta}}\right)\ket{\psi}\neq0$, so that $\ket{\psi'}$ in \Cref{psi'_to_rule_out_one_blockage} is normalizable.\raggedright}, it is possible to construct a new $\ket{\psi'}$ for which it does.
We can achieve this through subtracting out the component of $\ket{\psi}$ parallel to $\ket{\omega_{p\perp\beta}}$, by defining
\begin{equation}\label{psi'_to_rule_out_one_blockage}
\ket{\psi'}=N \left(\mathds{1}-\ketbra{\omega_{p\perp\beta}}{\omega_{p\perp\beta}}\right)\ket{\psi},
\end{equation}
where $N$ is a normalisation constant. 
The equality $\braket{\omega_{p\perp\beta}}{\psi'}=0$ for a $\ket{\psi'}$ constructed in this manner follows, as the projection $\mathds{1}-\ketbra{\omega_{p\perp\beta}}{\omega_{p\perp\beta}}$ right-annihilates $\bra{\omega_{p\perp_\beta}}$.

Ref.~\cite{hance_counterfactuality_2024} studied multi-path interferometers with the goal of positively identifying the presence of a blockage (rather than as here, determining the location of one already presumed to exist). There it was useful to expand $P(\omega_p|\beta)$ in the form
\begin{equation}\label{KD_expansion}
P(\omega_p|\beta)=
P(\omega_p|\text{no blockage})
-2\text{KD}
+\text{EV}
\end{equation}
where
\begin{align}
P(\omega_p|\text{no blockage}&)=|\braket{\omega_p}{\psi}|^2,\label{P_given_no_blockage}\\
\text{KD}&:=\text{Re}\left[\braket{\omega_p}{\beta}\braket{\beta}{\psi}\braket{\psi}{\omega_p}\right],\label{KDTerm}\\
\text{EV}&:=|\braket{\omega_p}{\beta}|^2|\braket{\omega_p}{\psi}|^2,\label{EV_term}
\end{align}
are respectively the probability of detector $p$ activating in the absence of a blockage, a Kirkwood-Dirac factor~\cite{Kirkwood1933KD,Dirac1945KDDistr,ArvidssonShukur2024KD}, and an Elitzur-Vaidman term~\cite{elitzur_quantum_1993}. In that paper it was argued that the greatest quantum counterfactual gain may be found by choosing a $\ket{\psi}$ corresponding to the largest possible negative value of factor KD. This maximises $P(\omega_p|\beta)$, whereas our present goal has been to search for $\ket{\psi}$ such that $P(\omega_p|\beta)=0$. The above method achieves this for $\ket{\psi'}$ as defined in \Cref{psi'_to_rule_out_one_blockage} by making each of \Cref{P_given_no_blockage,KDTerm,EV_term} individually equal to $\braket{\omega_p}{\psi'}$.

An obvious follow-up question is whether it is possible to construct a state $\ket{\psi}$ such that if detector $\omega_p$ fires, multiple blockage locations may be ruled out. This may be particularly useful if a point in a game has been reached where there are only a few possible locations remaining, so that these may be targeted. For this to be the case for a set of blockage locations $\beta_1,...,\beta_l$, a $\ket{\psi}$ must be found such that 
\begin{equation}\label{simultaneous_blockage_equations}
\left\{
\begin{matrix}
\braket{\omega_{p\perp\beta_1}}{\psi}&=0\\
&\vdots \\
\braket{\omega_{p\perp\beta_l}}{\psi}&=0
\end{matrix}
\right. .
\end{equation}
As above, this can only be true if $\ket{\omega_{p\perp\beta_1}},...,\ket{\omega_{p\perp\beta_l}}$ span at most a $k\leq d-1$ dimensional subspace. 
If this is true, and the player selects a linearly independent subset of these,
\begin{equation}
\left\{v_1,...,v_k\right\}\subseteq\left\{\ket{\omega_{p\perp\beta_1}},...,\ket{\omega_{p\perp\beta_l}}\right\},
\end{equation}
then Procedure~\ref{gram_schmidt_procedure} may be used to find such an input photon state.

\subsection{Quantum Ruleset: Why choosing \texorpdfstring{$\ket{\psi}=\ket{F}$}{initial state F} makes a good first turn for Hofmann's interferometer}\label{why_F_is_optimal}

As the outcome $\omega_0,...,\omega_p$ of any turn is probabilistic, the player cannot in general guarantee they will rule out a given number of blockages.
They must also take into consideration the likelihood of a given outcome occurring,
\begin{equation}
P(\omega)=\sum_\beta P(\omega|\beta)P(\beta),
\end{equation}
and as priors $P(\beta)$ are updated after each turn, what is considered a good move will change as the game progresses. For the first turn however, where the priors are equal, we may make some strong conclusions on a good choice of input state.

In addition to all $P(\omega|\beta)$, for input states $\ket{\psi}=\ket{\beta}$ for all $\beta\in\mathcal{B}$, and for $\ket{\psi}=\ket{\omega_1},...,\ket{\omega_d}$, the information tables in Appendix~\ref{info_tables} list $(1/7)\sum_\beta P(\omega|\beta)$. This is the overall probability of each outcome $P(\omega)$ on the first turn. For reference, the blockage probabilities $P(\omega|\beta)$ for $\ket{\psi}=\ket{F}$ are also given in \Cref{tab:blockedstateF}.

\begin{table}[h]
    \centering
\begin{tabular}{|c|ccccccc|c|}
\hline
\multicolumn{9}{|c|}{$P(\omega|\beta)$ for $\ket{\psi}=\ket{F}$}\\
\hline
&
$D_1$&
$S_1$&
$P_1$&
$F$&
$P_2$&
$S_2$&
$D_2$&
$\sum/7$
\\
\hline
 $\omega_0$&2/3 & 0 & 0 & 1 & 0 & 0 & 2/3 & 1/3 \\
 $\omega_1$&0 & 1/3 & 1/3 & 0 & 1/3 & 1/3 & 1/3 & 5/21 \\
 $\omega_2$&0 & 1/3 & 1/3 & 0 & 1/3 & 1/3 & 0 & 4/21 \\
 $\omega_3$&1/3 & 1/3 & 1/3 & 0 & 1/3 & 1/3 & 0 & 5/21 \\
 \hline
\end{tabular}
 \caption{Conditional probabilities $P(\omega|\beta)$, for Hofmann's three-path interferometer, of turn outcomes $\omega_0$--$\omega_3$ for input state $\ket{\psi}=\ket{F}$, and blockage locations $\beta$ given in columns.}
    \label{tab:blockedstateF}
\end{table}

The number of blockages ruled out by each outcome may be found by counting the number of zeros in each row. Evidently, given an input state $\ket{F}$ is absorbed (outcome $\omega_0$), the player would be able to rule out four of the seven locations. This property is not unique to input state $\ket{F}$; input states equal to $\ket{S_1}, \ket{S_2}, \ket{\omega_1}$, and $\ket{\omega_3}$ all have the potential to rule out four locations should a corresponding outcome be observed. The key reason $\ket{\psi}=\ket{F}$ is a better choice for the first input state is that, on the first turn, each of these other input states has only around a 5\% chance of producing the correct outcome to rule out 4 states, while $P(\omega_0)$ given input $\ket{F}$ is 33.3\%. Moreover, the fewest number of locations ruled out given first input state $\ket{F}$ is two (through outcomes $\omega_1$ and $\omega_3$), and there is also a 19.0\% chance of ruling out three locations (through outcome $\omega_2$). The particularly large number of zeros in the state $\ket{F}$ table singles it out among the other special states considered as a good first turn.\footnote{$\ket{\psi}=\ket{F}$ also has the best expected information gain we have found for a starting move---see \Cref{info_gain_optimization,mixed_state_info_gain}.}

In a finite number of turns, it \emph{is possible} to finish a game using state $\ket{F}$ alone, but only if the blockage is in location $D_1$ or $D_2$. To see this, first observe that to complete a game solely using $\ket{\psi}=\ket{F}$, the player must, in a sequence of turns, observe a set of outcomes that rules out all but one blockage. Hence there must exist a subset of rows in which all but one of the the corresponding columns contain a zero. This cannot be the set of all rows (outcomes) as there is at least a single zero in every column (blockage) of the table, so that if in a sequence of turns all outcomes occur, then all blockages would be ruled out, which clearly cannot happen. (Indeed the same may be said of outcomes $\omega_0$ and $\omega_2$; in a single game it is impossible to observe both outcomes when using input state $\ket{F}$.) However, for all columns (i.e., all possible blockage locations), the sub-columns corresponding to $\{\omega_0,\omega_1\}$ contain zeros except for $D_2$. Therefore, if in the first two turns of a game the player uses state $\ket{F}$, and obtains outcomes $\omega_0$ and $\omega_1$, they may conclude the game having determined the blockage to be at position $D_2$. Similarly, if they obtain outcomes $\omega_0$ and $\omega_3$, they may conclude the blockage is at $D_1$.

Of course this is a property of only $D_1$ and $D_2$. If instead $\omega_2$ is observed, then $D_1$, $F$, and $D_2$ are ruled out, but as the remaining columns are identical, nothing more may be learned from inputting exclusively state $\ket{F}$; state $\ket{F}$ cannot distinguish between states $S_1$, $P_1$, $P_2$, and $S_2$. If, on the other hand, the blockage is at position $F$, then only outcome $\omega_0$ may be recorded, but the player can never be fully certain the photon was blocked at position $F$. After the first turn the players priors are updated to $P(\beta=F)=3/7$ and $P(\beta=D_1)=P(\beta=D_2)=2/7$, with all others vanishing. After that $P(\beta=F)$ follows the recurrence relation
\begin{equation}
P(\beta=F)\to\frac{3P(\beta=F)}{2+P(\beta
=F)},
\end{equation}
which, denoting the probability after the $n^\text{th}$ turn $P^n$ has solution $P^n(\beta=F)=3^n/(2^{n+1}+3^n)$ for $n\geq1$. It takes fourteen turns for this to exceed 99\%.

\subsection{Both Rulesets: A strategy of information gain maximisation}\label{info_gain_optimization}
Rather than attempting to rule out blockages, a player may instead aim to optimise the reduction in entropy $H(B)$, or equivalently maximise information gain $I(B;\Omega=\omega)$.
As ever, the probabilistic outcomes $\omega$ make this an uncertain process.
A straightforward strategy to attempt to maximise information gain $I(B;\Omega=\omega)$ is to choose a $\ket{\psi}$ that maximises its expectation value,
\begin{equation}
I(B;\Omega):=
\langle I(B;\Omega\mathord{=}\omega)\rangle
=\sum_{\omega'\in\mathcal{O}}P(\omega')I(B;\Omega\mathord{=}\omega').
\end{equation}
A player who is able to choose the optimal $\ket{\psi}$ to maximise $I(B;\Omega)$, must still react to the probabilistically evolving $P(\omega)$, so that optimal input photon states and game lengths will vary across games. 
On each turn, the player may, for any potential input state $\ket{\psi}$, calculate the expected information gain using the routine described in \Cref{player_data_gathering}.

\begin{algorithm}[H]
  \caption{Example player data gathering routine.}
  \label{player_data_gathering}
\parbox{0.90\linewidth}{
   \begin{algorithmic}[1]
\State set $\Entropy\leftarrow -\sum_\beta\Priors(\beta)\log\Priors(\beta)$
\State select $\TrialInputState$
\State for all $\beta,\omega$ use this to calculate $\ProbDgB(\omega,\beta)$
\Statex \Comment{for instance by using \Cref{quantum_forward_probs} or \Cref{classical_forward_probs}\hspace{5ex}}
\State for all $\omega$ set
\begin{align*}
\ProbD(\omega)\leftarrow \sum_{\beta}\Priors(\beta)\ProbDgB(\omega,\beta)
\end{align*}
\State for all $\beta,\omega$ set
\begin{align*}
\ProbBgD(\beta,\omega)\leftarrow \frac{\Priors(\beta)\ProbDgB(\omega,\beta)}{\ProbD(\omega)}
\end{align*}
\State for all $\omega$ set
\begin{align*}
\PotEntropy(\omega)\leftarrow -\sum_\beta\ProbBgD(\beta,\omega)\log\ProbBgD(\beta,\omega)
\end{align*}
\State set $\ExpectedEntropy\leftarrow \sum_\omega\ProbD(\omega)\PotEntropy(\omega)$
\State set $\ExpectedInfoGain \leftarrow \Entropy-\ExpectedEntropy$
   \end{algorithmic}
}
\end{algorithm}

A choice of optimisation algorithm may be employed to iteratively use \Cref{player_data_gathering} to locate an input state with maximal expected information gain. For the below numerical results we use a simple brute force Monte Carlo method, generating many random quantum states and selecting that with the largest $I(B;\Omega)$. 
However, this still leaves open the question of how the random quantum states are sampled.
Interestingly, in our numerical simulations of games, an information gain optimizing algorithm that samples the space of pure states (the projective Hilbert space) uniformly, fails to finish games. 
Such an algorithm performs well in the early game, reducing uncertainty quickly, and tends seemingly arbitrarily close to $H(B)=0$, but never actually reaches $H(B)=0$. 
We believe the reason for this is that, in order to complete a game, a sequence of turns must take place in which all but one blockage location is ruled out. That is, for some turns $\ket{\psi}$ must be chosen such that $P(\beta|\omega)=0$ exactly, which as we have discussed above requires $P(\omega|\beta)=0$.
The conditions for this are $\braket{\beta}{\psi}=0$ for outcome $\omega_0$, and $\braket{\omega_{p\perp\beta}}{\psi}=0$ for outcomes $\omega_p$ with $p\geq 1$.
For either of these conditions to hold, $\ket{\psi}$ must live on a measure zero subspace of the projective Hilbert space, and so if this space is uniformly sampled, a state satisfying either of these criteria will be selected with probability zero. 

This problem can of course be rectified with better sampling (or with a better optimization algorithm).
In \Cref{sec:results} we compare information gain optimized uniform sampling to that of both sampling of only the blockage vectors, and of the combination of both. We also apply the information gain optimization to the classical ruleset.

\subsection{Quantum Ruleset: Mixed states}

\begin{figure}
\includegraphics[width=\columnwidth]{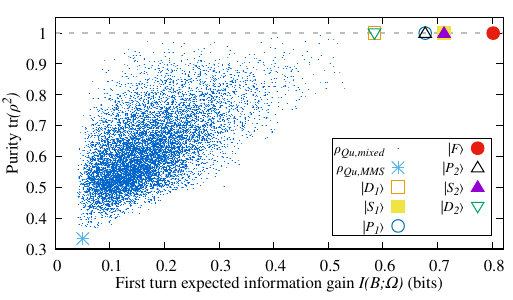}
\caption{For mixed states, the first turn expected information gain $I(B;\Omega)$ is strongly correlated with purity. The maximally mixed state $\rho_\text{Qu,MMS}$ is particularly low in this regard, though not quite the lowest. Pure states $\ket{\psi}=\ket{\beta}$ are shown for comparison.}\label{mixed_state_info_gain}
\end{figure}

For density matrices $\rho_\text{Qu}$ of a range of different purity, \Cref{mixed_state_info_gain} shows the first turn estimated information gain $I(B,\Omega)$ in bits, using Hofmann's three-path interferometer. As expected, states of higher purity generally give greater $I(B,\Omega)$ than states of lower purity, with the seven possible blockage location states giving the highest possible $I(B,\Omega)$ of the states chosen. The maximally mixed state $\rho_\text{Qu,MMS} = \sum_{i=1}^3\ketbra{i}{i}$ gives far lower $I(B,\Omega)$, but perhaps surprisingly isn't the worst state in this regard. Inspecting the output probabilities for $\rho_\text{Qu,MMS}$ for various blockage locations (\Cref{table:maxmixedprobs}), we see that all blockages result in a 33\% chance of absorption. This means we learn absolutely nothing for this state if the photon is absorbed: all information gain for this state must come from cases where the photon reaches one of the detectors. Also interestingly, $\rho_\text{Qu,MMS}$ gives the same output probabilities for a blockage at $D_1$ and $S_1$, and for $D_2 $ and $S_2$, and is thus is unable to distinguish between these states.
\begin{table}
\begin{tabular}{|c|ccccccc|c|}
\hline
&
$D_1$&
$S_1$&
$P_1$&
$F$&
$P_2$&
$S_2$&
$D_2$&
$\sum/7$
\\
\hline
 $\omega_0$&0.333 & 0.333 & 0.333 & 0.333 & 0.333 & 0.333 & 0.333 & 0.333 \\
 $\omega_1$&0.167 & 0.167 & 0.278 & 0.222 & 0.111 & 0.333 & 0.333 & 0.230 \\
 $\omega_2$&0.167 & 0.167 & 0.278 & 0.222 & 0.278 & 0.167 & 0.167 & 0.206 \\
 $\omega_3$&0.333 & 0.333 & 0.111 & 0.222 & 0.278 & 0.167 & 0.167 & 0.230 \\
 \hline
\end{tabular}
\caption{$P(\omega|\beta)$ for $\rho_\text{Qu,MMS}$ and blockage indicated in columns. Note no entry is zero.}\label{table:maxmixedprobs}
\end{table}

To help understand why this is the case consider that, given a mixture
\begin{equation}
\begin{split}
\rho_\text{Qu}&=\sum_ip_i\ketbra{\psi_i}{\psi_i}, \quad p_i > 0,\quad \sum_i p_i=1,
\end{split}
\end{equation}
conditional probabilities are merely
\begin{equation}\label{mixed_forward_conditional_detected}
P(\omega_p|\beta)=\sum_ip_iP_i(\omega_p|\beta)
=\sum_ip_i|\braket{\omega_{p\perp\beta}}{\psi_i}|^2,
\end{equation}
where $P_i(\omega_p|\beta)$ is the conditional probability given pure state $\ketbra{\psi_i}{\psi_i}$. Therefore, for \Cref{mixed_forward_conditional_detected} to vanish, so that blockage $\beta$ may be ruled out, \Cref{forward_prob_zero_detected} would need to hold for all components in the mixture. Similarly,
\begin{equation}\label{this_other_equation}
0=P(\omega_0|\beta)
=1-\sum_i\sum_{p=1}^d p_iP_i(\omega_p|\beta)
=\sum_ip_i|\braket{\psi_i}{\beta}|^2
\end{equation}
only if $|\braket{\psi_i}{\beta}|^2=0$ for all $i$.

Rather than viewing $\rho_\text{Qu}$ as a mixture of pure states, we may instead choose to view it as an abstract density matrix. The condition $P(\omega_p|\beta)=0$ may then be expressed
\begin{equation}\label{this_equation}
0=\Tr\left[\rho_\text{Qu}\ketbra{\omega_{p\perp\beta}}{\omega_{p\perp\beta}}\right].
\end{equation}
Expanding the density out in its spectral decomposition $\rho_\text{Qu}=\sum_i\lambda_i\ketbra{v_i}{v_i}$, we see for \Cref{this_equation} to hold, that
\begin{equation}
0=\sum_i\lambda_i|\braket{\nu_i}{\omega_{p\perp\beta}}|^2.
\end{equation}
Thus for all $i$, either $\lambda_i=0$ or $\braket{\nu_i}{\omega_{p\perp\beta}}=0$ (or both). This condition is the eigenbasis equivalent to \Cref{mixed_forward_conditional_detected}. Equivalently, $\rho_\text{Qu}\ketbra{\omega_{p\perp\beta}}{\omega_{p\perp\beta}}=0$. Finally to rule out a blockage when the photon is blocked, we require
\begin{equation}
0=P(\omega_0|\beta)
=\bra{\beta}\rho_\text{Qu}\ket{\beta}
=\sum_i\lambda_i|\braket{\beta}{v_i}|^2,
\end{equation}
which is the eigenbasis equivalent to \Cref{this_other_equation}.

\begin{figure}
\includegraphics[width=0.85\linewidth]{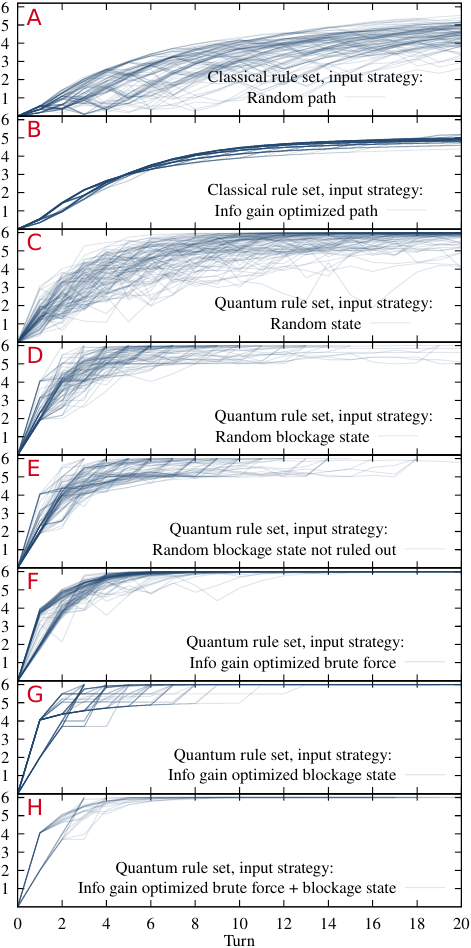}
\caption{Comparison of game progress using two strategies in the classical ruleset and six strategies in the quantum ruleset. 
Each frame shows 100 games played with the indicated strategy and ruleset.
For consistency, the blockage location is kept at $D_1$.
Equivalent strategies with the quantum ruleset clearly outperform their classical analogues. 
Turns that result in a negative information gain are seen as temporary downward movements, with these more common for strategies that do not optimize information gain. 
Brute force information gain optimisation entails choosing the state with maximum estimated information gain from a sample of 1000 uniformly sampled input states at each turn.
This performs comparatively well in the early game, and tends towards 6 blockages ruled out, but due to the sampling does not quite manage to finish the game completely (see \Cref{info_gain_optimization}). 
Perhaps unsurprisingly, the best performer is the information gain algorithm with the best state sampling.
}\label{fig:comparison}
\end{figure}

\section{Discussion}\label{sec:Discussion}
\subsection{Preliminary numerical comparison of strategy performance}\label{sec:results}
A comprehensive numerical evaluation of strategy performance would need to include both a variety of interferometer configurations and sizes, and more sophisticated sampling techniques than we have so far programmed, and so would constitute a substantial extra project.
Nevertheless, our initial analysis\footnote{See Appendix \ref{coding_guidance} for additional guidance on numerical simulation of games.} for Hofmann's three-path interferometer sheds some light on the relative performance of strategies, confirming some of our analytic assertions as well as highlighting some unexpected properties.

Each frame in \Cref{fig:comparison} shows the turn-by-turn evolution of the equivalent number of blockages ruled out, $7-2^{H^{(n)}}$, for 100 games played with a particular strategy.
For consistency, the blockage location is kept as $D_1$.
Frames A and B are strategies played with the classical ruleset, and frames C to H are strategies played with the quantum ruleset.
Baseline ``control group'' strategies, played with random input states, are shown for comparison in frames A and C. 
These do make game progress (as measured by entropy/equivalent blockages ruled out), demonstrating that random data is better than no data, though progress is slower than for the non-control strategies.
No member of these control groups manages to finish a game, and locate the blockage.
For the classical ruleset perfectly locating the blockage is typically impossible, as may be seen through inspection of \Cref{table:classical_probs}, which contains $P(\omega|\beta)$ for the classical ruleset.
As discussed in \Cref{ruling_out_blockages}, Bayes' theorem tells us that to obtain a posterior $P(\beta|\omega)$ equal to zero for a non-zero prior $P(\beta)$, and thus rule out a blockage, we require $P(\omega|\beta)=0$.
Hence \Cref{table:classical_probs} tells us that only blockage locations $D_1$ and $S_1$ may be ruled out with the classical ruleset. 
For the quantum ruleset control group, progress in ``equivalent blockages ruled out'' is significantly quicker, though for the same reason discussed in \Cref{info_gain_optimization}, due to the uniform sampling of the space of pure states (upon which subspaces of states capable of ruling out blockage locations are measure zero) this group fails to rule out blockages and complete a game.
That said, zero entropy is approached (at least by eye) asymptotically.

\subsection{Classical vs Quantum}\label{sec:classical_vs_quantum}

Notably in \mbox{\Cref{fig:comparison}} we can see that not only does random classical path (A) perform worse than nearly all quantum games (C-H), but information gain-optimised classical path (B) similarly performs worse than nearly all quantum games, even quantum random state (C). We can interpret this by considering the different role beam splitters play for the classical game compared to the quantum game. In the quantum game, beam splitters move us between contexts: the state after the beamsplitter remains the same as the state before the beamsplitter, just expressed in a different basis. However, in the classical game, beamsplitters insert probabilistic behaviour: unless they all have either reflectivity 0 or reflectivity 1, there is no sequence of beam splitters for the classical game which will cause the ``photon'' to deterministically arrive at a certain location. Interference, which in e.g., Hofmann's interferometer ensures for the quantum ruleset that, if there are no blockages, a photon input into $\alpha_2$ will always end up at $\omega_2$, does not exist for the classical ruleset. Due to interference, we can view the quantum ruleset as preserving information across beamsplitters, first about the initial state and then through how the state is altered the blockage location. Beam splitters in the classical ruleset seem instead to ``wash away'' this information: successive probabilistic reflections/transmissions before the blockage progressively remove any information the ``photon'' carries about its input path, then successive probabilistic reflections/transmissions after the blockage progressively remove any information the photon carries about the blockage location (or rather, the path the ``photon'' took to bypass the blockage.

While nowhere near as easy to formalise through statistical distances into a ``quantum counterfactual gain'', in the same way as we did in Ref.~\mbox{\cite{hance_counterfactuality_2024}}, we hope future work will allow us to use this difference between how beam splitters manifest in the two rulesets to more formally identify the advantage quantum resources provide for the game.

\subsection{Benefits of Losing Information}

The effect of imposing information gain optimisation on the classical ruleset may be seen in frame B, which displays a quicker and more reliable rise in effective blockages ruled out than control group A.
The equivalent quantum effect is seen by comparing frames C and F, each of which sample the space of pure states uniformly.
The information gain optimisation algorithm employed in frame F (the simple brute-force Monte Carlo scheme, sampled over the projective Hilbert space, mentioned earlier) specifically targets entropy reduction/equivalent blockages ruled out, and so it is unsurprising that this outperforms the control group (while still failing to finish a game).

Quantum ruleset strategies for which the sampling is reduced to only the set of blockage vectors, so that $\ket{\psi}=\ket{\beta}$ for some blockage $\beta$, are shown in frames D,E, and G.
Respectively these strategies select a random blockage state (serving as a control for this sub-group), a random state not ruled out (which benefits from the contrapositional logic of Strategy~\ref{simple_strategy}, and so outperforms the control group), and an expected information gain optimized blockage state.
The last of these performs particularly well, and is only marginally outperformed by frame H.
Frame H is a ``best of all worlds'' strategy, where information gain optimisation is used, but the sampling combines the brute force uniformly sample pure states and the blockage vectors.
The difference between this and frame G indicates that (at least as measured by expected information gain), a blockage state is not always optimal.

\begin{figure}
\centering
\begin{subfigure}{\linewidth}
    \captionsetup{ justification=raggedright, singlelinecheck=false }
    \caption{}\label{fig:3d_histograms_subfig_randomblockage}
		\includegraphics[width=\textwidth]{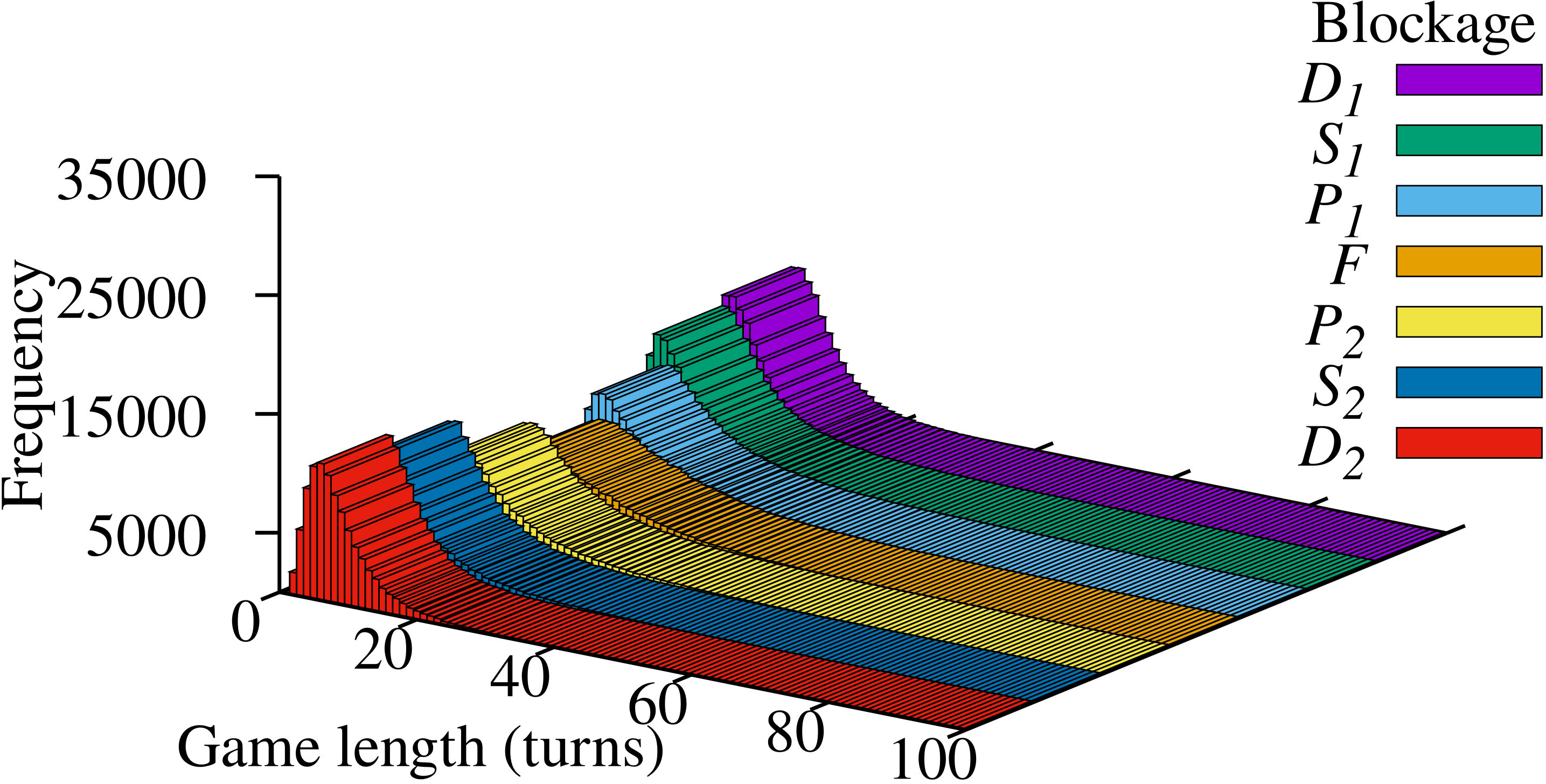}
	\end{subfigure}

\begin{subfigure}{\linewidth}
    \captionsetup{ justification=raggedright, singlelinecheck=false }
    \caption{}\label{fig:3d_histograms_info_opt_blockage}
		\includegraphics[width=\textwidth]{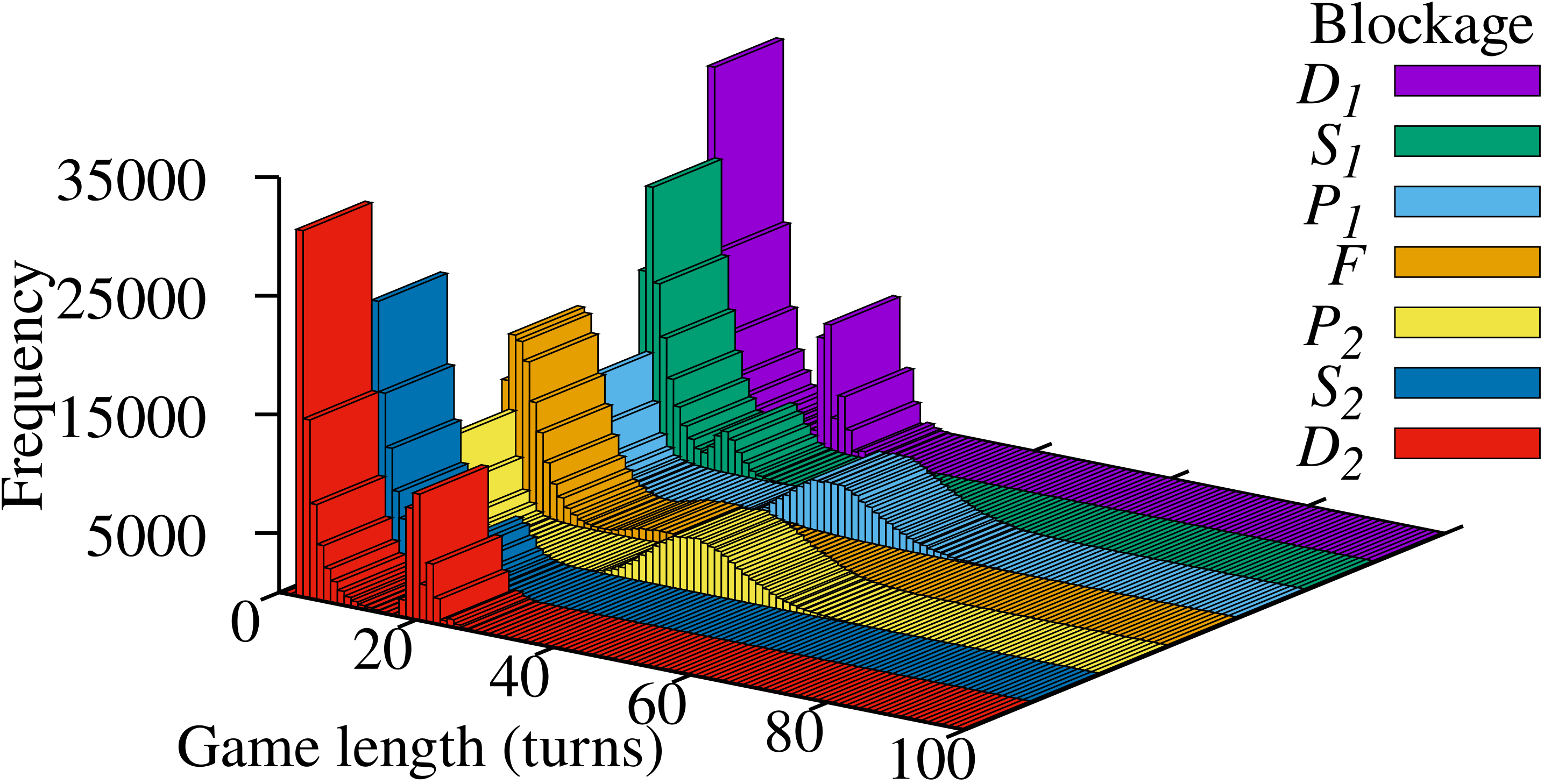}
	\end{subfigure}
\caption{Histograms of game duration for 100,000 games with each blockage, played using: In \Cref{fig:3d_histograms_subfig_randomblockage} input photon states $\ket{\psi}=\ket{\beta}$ for a random blockage $\beta$ yet to be ruled out, and in \Cref{fig:3d_histograms_info_opt_blockage} $\ket{\psi}$ equal to the $\ket{\beta}$ with the best expected information gain.
The densities of information optimized game length are bimodal; in the majority of cases they outperform the random choice, however for a smaller population of games they underperform.
A loose analogy may be taken with ``risky'' strategies in expected utility theory. (See discussion in \Cref{sec:results}).}\label{fig:3d_histograms}
\end{figure}

So far, the discussion may suggest that information gain optimisation (as we have described it) is always beneficial, but we have found one major drawback.
We illustrate this in \Cref{fig:3d_histograms}, which compares histograms of game lengths for a strategy of selecting input states $\ket{\psi}$ equal to $\ket{\beta}$ for blockages $\beta$ yet to be ruled out (\Cref{fig:3d_histograms_subfig_randomblockage}),  with that of $\ket{\psi}$ equal to $\ket{\beta}$ for the $\ket{\beta}$ with the maximum expected information gain (\Cref{fig:3d_histograms_info_opt_blockage}).
In the majority of games, and for all blockage locations, information gain outperforms the simpler strategy.
However, each histogram in \Cref{fig:3d_histograms_info_opt_blockage} is conspicuously bimodal, with a second smaller but not insignificant population of games that underperform compared to the random strategy.
Moreover, for every blockage location, the two populations (underperforming and overperforming) appear fully disjoint, which we think is suggestive of a single failure mode for the information gain approach.

At present, we are yet to pin down the mechanism for this suspected failure mode although we believe it could be due to an over-aversion to ``losing information''.
Maximisation of expected information gain makes it probable that entropy will be reduced, while at the same time making it unlikely that entropy will increase, and so it encourages further peaking the density of priors around the blockage where it is already peaked.
However, unlikely outcomes in the early game may mean that this peak may begin around the wrong location. 
If this happens, then the algorithm may have to, in a manner of speaking, unlearn the lessons it learned through these early unlikely events.  
This will include lowering of value of the previously-considered most likely location, and may increase entropy overall. 
The algorithms used in \Cref{fig:3d_histograms_subfig_randomblockage,fig:3d_histograms_info_opt_blockage} are displayed in a turn-by-turn fashion for individual games in frames E and G of \Cref{fig:comparison}.
Here we note that the information gain optimisation strategy doesn't appear to ever lose information (as would be seen by a downtick in the curve), whereas this appears a quite common occurrence for the random strategy.
Of course at present this is mere speculation, and would certainly require  more investigation to substantiate.

The potential for strategies of decision making under uncertainty to fail is not unknown.
In microeconomics, similar scenarios are widely studied under the banner of Expected Utility Theory~\cite{sep-rationality-normative-utility}.
Crudely speaking, there the maximisation of expected ``gain'' is often considered risk-neutral, and to deviate from this neutrality one can apply to the gain either a concave ``utility'' function (to lean towards risk aversion), or a convex utility function to lean towards more risk.
There are clearly important differences between that case and this, however the analogy at least points to the potential value of the application of a utility function, and by extension the attention from experts in game theory.

\subsection{Non-demolition blockages}\label{sec:ndspd}

We can view the Quantum Plumber problem as equivalent to there being a truncation of the Hilbert space from $d$ to $d-1$ dimensions, where we just wanting to find out which dimension was the one truncated. As a future extension, we could instead consider the situation where the block is a non-demolition measurement (i.e., NDSPD~\cite{Braginsky1980QND,Kok2002LinearOpticQND,Besse2018PhotonQND}), which instead bifurcates the Hilbert space (albeit with probabilities at $\omega_i$s being sums of probabilities from the two bifurcations). It is interesting to consider how much harder the game would be for both a classical and a quantum player in these circumstances. Given a non-demolition detection removes coherence but not probability, in this case a classical player would have just random luck chance of guessing right, meaning the bar for a quantum player to get an above-classical performance would be lower.

A non-demolition detector at location $\beta$ whose result is not reported to the player replaces \mbox{\Cref{quantum_forward_probs_pure_state}} with
\begin{equation}\label{ndspd_probs}
 P(\omega_p|\beta)=|\braket{\omega_p}{\beta}|^2|\braket{\beta}{\psi}|^2+\left|\bra{\omega_p}P_{\perp\beta}\ket{\psi}\right|^2
\end{equation}
for $p=1,...,d$, with $P(\omega_0|\beta)=0$. Under the classical ruleset the detector changes nothing, so $P(\omega|\beta)$ is independent of $\beta$ and the classical player can never update their priors. Under the quantum ruleset, a location can be ruled out on outcome $\omega_p$ only if both terms in \mbox{\Cref{ndspd_probs}} vanish, which is why we included the detector states $\ket{\omega_p}$ in the set $V$ of \mbox{\Cref{info_gain_optimization}}. For Hofmann's interferometer there is a further restriction: the pair $D_1,S_1$ gives identical conditional probabilities for every input state, as does the pair $S_2,D_2$. Both members of each pair lie in the plane spanned by two of the detector states, so the two dephasing operations differ only in coherences that the final measurement cannot see. We would therefore need to score this variant by the number of turns needed to concentrate the posterior on one of the five distinguishable classes $\{D_1,S_1\},P_1,F,P_2,\{S_2,D_2\}$, with the uniform prior over the seven locations. This variant is therefore harder for the quantum player, but for the classical player cannot be completed at all.

\subsection{Future Work}

As discussed briefly at the end of \mbox{\Cref{sec:results}}, it would be interesting to consider more ``risk-prone'' or ``risk-adverse'' strategies for information gain, where we aim not to maximise the expectation value of the information gain itself, but the expectation of some function of the information gain. We will consider this further in future work.

One benefit of the single-photon/``classical photon'' nature of the game common to both rulesets is that each turn, there is only one outcome $\omega_0-\omega_d$, and these outcomes are mutually exclusive: we know, for instance, in turns where the photon arrived at $\omega_2$, that it didn't also arrive at (and so wasn't also absorbed by) the blocker. This discreteness however is only a property of very specific quantum states of light (the Fock, or number, states), of which we only look at one sub-example (the single-photon state). This suggests both a narrow and a broader path for potential expansion of the game. The narrow expansion would be to consider other number states (i.e., cases where more than one indistinguishable photon was input into the interferometer). Such an expansion would link the game to boson sampling~\mbox{\cite{Brod2019BosonSampling}}, albeit one with the additional caveat of potential photon loss. It also would mean the output detections would no longer be mutually-exclusive---for an $N$-photon input state, we would get $N$ outputs, each sampled from $\omega_0-\omega_d$. Alongside this, such an extension would introduce novel multi-photon effects without any single-photon analogue, like the Hong-Ou-Mandel effect~\mbox{\cite{Hance_2025}}, or wider bunching/anti-bunching effects, which may have additional interesting informational benefits (or drawbacks) for the game. The broader expansion would be to consider the wider range of non-Fock states of light, such as coherent states, Gaussian states, thermal states, or even novel engineered states previously used for metrology. As with $N>1$ Fock states, these again have the issue of making the output detections no longer mutually exclusive; however, unlike the Fock states, we can't even bound the number of outputs we would see, given all these states are superpositions of different photon-number states, so give a probability distribution over different photon numbers when measured in the photon-number basis (as we do when measuring with single-photon detectors, like here). Despite this, for certain states (e.g., coherent states), the overall distribution of the ``clicks'' at different outputs should match the distribution for the single-photon case---the only difference would be that for a given input state we wouldn't know (beyond an expectation value) in advance how many times we would be sampling from this distribution for the outputs. 

This last point also affects \mbox{\Cref{sec:classical_vs_quantum}}'s comparison: a coherent state is a classical wave, and it reproduces the single-photon conditional probabilities per unit intensity, so the advantage of the quantum ruleset over the pinball ruleset is an advantage of interference over particle-like propagation rather than something that requires a single-photon source. Single photons add the mutual exclusivity of the outcomes: quantifying the advantage gained through this is a question for the multi-photon extension. Relatedly, the optimal states for estimating the magnitude of a loss in a bosonic channel are known to be Fock states~\mbox{\cite{Monras2007Loss,Adesso2009Loss}}, and the extension of the game to lossy components (as discussed in \Cref{sec:interferometer}) would turn the discrete problem considered here into a problem of that type.

A feasible direction for future work would be to link our analysis to recent work by Hance \emph{et al}~\cite{hance2026noncontextualversuscontextualinterferometry} showing the extent to which such a three-path interferometer can and cannot be represented using either Quantum Simulation Logic (based on Spekkens' toy model) or the stabiliser subset of quantum mechanics, and use this to link advantage in this game to ``magic'' (or ``non-stabiliser-ness'') in the interferometer and initial state.

An additional idea for future work is a version of the Quantum Plumber's problem where the ``cost'' which we want to minimise is the photon being lost, rather than the turns taken - for this we would consider detectors $\omega_i$ for $i\neq 0$ (i.e., $\omega_1$--$\omega_3$ for Hofmann's interferometer) as being non-demolition, so the photon is not lost if it goes to these detectors, but where the photon is still lost for $\omega_0$. We can imagine, for such a game, starting with $x$ photons to use - once you're done, you're done. The question is, under these constraints, how does the optimal strategy change. This is the constraint imposed in Ref.~\mbox{\cite{Chaturvedi2026IFLocalization}}, where the one-shot optimum and a finite-round adaptive benchmark are derived for a designable multiport interferometer, and it is closely related to absorption-free discrimination between semi-transparent objects~\mbox{\cite{Mitchison2001AbsorptionFree,Massar2001MinimalAbsorption}} and to the quantum Zeno schemes that make the Elitzur-Vaidman test efficient~\mbox{\cite{Kwiat1995IFM,Kwiat1999Zeno,Zhou2017IFMChannel}}.

Yet another interesting direction would be to remove the exact limit of there being a single blockage, and instead to have each location having its own (i.i.d.) probability of being blocked. Although in Hofmann's interferometer as considered, each blockage has a probability of 1/7, if this was 1/7 i.i.d, then there would be a possibility of 0 blockages, and a possibility of $>1$ blockage, affecting our analysis here.

Beyond this, an avenue for future work is the ``semi-quantum'' approach of being limited to input states $\alpha_1,\alpha_2,\alpha_3$, rather than arbitrary superpositions of these. For Hofmann's interferometer, we can see how this would affect the game from the tables for $\ket{\omega_1}$, $\ket{\omega_2}$, and $\ket{\omega_3}$ in Appendix~\mbox{\ref{info_tables}}, since the input and output port states coincide: none of the columns $P_1$, $F$, and $P_2$ contains a zero, and the columns $D_1$ and $S_1$ (and $S_2$ and $D_2$) are identical. A player restricted to the input ports can rule out $D_1$ and $S_1$ together, or $S_2$ and $D_2$ together, but can never rule out $P_1$, $F$, or $P_2$, and so cannot complete the game. For other interferometers the restricted game may be completable, but this remains an open question.

Given we can view the game as a sequential decision-making problem, there is also the question of whether we could tie a potential strategy to reinforcement learning, where we try to learn an optimal policy for what state we should send in our next photon, based on the outcome of the previous ones. Since the priors $P(\beta)$ summarise everything the player has learned, the game is a partially observable Markov decision process whose belief state is the vector of priors~\mbox{\cite{Kaelbling1998POMDP}}, and the objection to tabulating optimal moves raised in \mbox{\Cref{sec:Game}} does not apply to a policy indexed by the belief state rather than by the history.

Finally, a blocked path is a marked item and each photon is one query, so the game is a form of search by interaction-free measurement, which has been used to implement Grover's algorithm counterfactually~\mbox{\cite{Hosten2006Counterfactual}}.

\subsection{Summary}

In this paper, we extended previous analysis of ``quantum counterfactual gain''~\mbox{\cite{hance_counterfactuality_2024}} to identifying quantum advantage in a new scenario, which we term the ``Quantum Plumber's Problem''.
In this scenario, which we represented as a game, we cast the player as a ``quantum plumber'', who knows that one path of an interferometer is blocked, and wants to find the optimal strategy for identifying, with certainty, \emph{which} path this is. We considered various strategies for a generalised path-encoded interferometer, as well as giving numerics for the specific case of Hofmann's three-path interferometer \mbox{\cite{hofmann_sequential_2023}}, which allowed us to build our intuition about the game for a simple but non-trivial concrete scenario. Hofmann's three-path interferometer allowed us to debate the relative merit of competing strategies with data collected over many simulated attempts at locating blockages.

 \textit{Acknowledgements---} We thank Sarah Croke, Jakob Dautovic Bergh, Donald Spector, and the participants of QIP 2026 V\"axj\"o for useful suggestions for extensions/future work. JRH acknowledges support from a Royal Society Research Grant (RG/R1/251590) and an EPSRC Mathematical Sciences Small Grant (UKRI3647). JRH and NGU acknowledge support from JRH's EPSRC Quantum Technologies Career Acceleration Fellowship (UKRI1217). HFH acknowledges support from ERATO, Japan Science and Technology Agency (JPMJER2402).

\bibliographystyle{unsrturl}
\bibliography{references.bib}

\begin{appendix}
\section{Practical guidance for numerical simulation of games}\label{coding_guidance}
While a numerical simulators of games with basic strategies are relatively straightforward to program, there are two potential pitfalls that we wish to note for researchers wishing to reproduce or extend our analysis.
Firstly, games conclude when $P(\beta)=\delta(\beta,\gamma)$, which is \emph{exactly} unity for $\beta=\gamma$ and \emph{exactly} zero for all $\beta\neq\gamma$.
Of course, the standard 64-bit floating point arithmetic used on the vast majority of computers rounds calculations at the $16^\text{th}$ significant figure, and so calculations of $P(\omega|\beta)$ that should result in zero analytically, rarely do numerically, most often under or over-shooting over-shoot (in the latter case producing negative probabilities).
To prevent this, it is necessary to introduce a cut-off to the calculations.
We have found that such calculations that should return zero analytically, usually return a value of magnitude $\sim10^{-12}-10^{-13}$ numerically.
To ensure these are instead evaluated as zero without creating other issues, we set any posterior probability $P(\beta|\omega)$ calculated to be less than $10^{-10}$ to zero, and renormalise the rest accordingly.
This cut-off has a side effect that should be kept in mind when reading game-length statistics: a strategy whose posterior converges geometrically towards a single location without ever ruling out the alternatives (see \mbox{\Cref{sec:results}}) will be recorded as having finished once the alternatives fall below the cut-off, at a turn number set by the cut-off rather than by the strategy.
(However, with the cut-off set as described, we have found this happens only for game lengths far longer than we have considered in our analysis.)

Second, the job of an information gain optimization algorithm is to select the state $\ket{\psi}$ sampled with the maximum expected information gain.
However, symmetry in the problem can result in two or more sampled states with the same maximal analytic information gain.
For instance, for the Hoffman interferometer (which has a mirror symmetry down a central vertical axis), take an information gain optimized choice of blockage vectors algorithm.
As we have discussed, the first turn will always be $\ket{\psi}=\ket{F}$.
If the outcome of this turn is $\omega_0$, then the second choice of input state becomes ambiguous.
Input state $\ket{\psi}=\ket{D_1}$ results in the same analytic expected information gain as $\ket{\psi}=\ket{D_2}$, and these are higher than any other value.
Numerically however, due to the order inherent in a computer program, one or the other will always be picked. 
This breaks the symmetry of the interferometer, biasing game length so that games are on average shorter for some blockages and longer for others. 
To address this, we have inserted a control statement into the information gain optimisation algorithm that, if it finds the most optimal states to have an expected information gain within $10^{-6}$ bits of each other, then chooses randomly between them.
We have found this to be sufficient to restore the symmetry of game length, so that for instance the histograms of games with blockages at $D_1$ and $D_2$, $S_1$ and $S_2$, and $P_1$ and $P_2$, in are the same in \Cref{fig:3d_histograms_info_opt_blockage}.

\begin{widetext}
    \section{Probability Tables for different Input States}

See \Cref{HofmannTablesFull} for tables giving $P(\omega|\beta)$ for a photon arriving at different detectors $\omega_1-\omega_3$ (or being absorbed $\omega_0)$, for different blocked paths $\beta$ as given by column headers, for different input states $\ket{\psi}$, for Hofmann's three-path interferometer~\cite{hofmann_sequential_2023}. Note there is a symmetry to these tables. For instance the $D_1$ and $D_2$ tables are the same with reversed columns if we also switch rows $\omega_1\leftrightarrow\omega_3$. The same is true for tables for input states $S_1$ and $S_2$, $P_1$ and $P_2$, and $\omega_1$ and $\omega_3$. This suggests that for instance games played with a blockage at $D_1$ should have the same statistics (average game length for instance) as games played with a blockage at $D_2$.

\label{info_tables}
\begin{table}[t]
\caption{Probability tables giving $P(\omega|\beta)$ for Hofmann's three-path interferometer for a photon arriving at different detectors $\omega_1-\omega_3$ (or being absorbed $\omega_0)$, for different blocked paths $\beta$ as given by column headers, for different input states $\ket{\psi}$.}\label{HofmannTablesFull}
\hfill
\begin{subtable}[t]{0.45\textwidth}
\begin{tabular}[t]{|c|ccccccc|c|}
\hline
\multicolumn{9}{|c|}{$\ket{\psi}=\ket{D_1}$}\\
\hline
&
$D_1$&
$S_1$&
$P_1$&
$F$&
$P_2$&
$S_2$&
$D_2$&
$\sum/7$
\\
\hline
$\omega_0$& 1.000 & 0 & 0.333 & 0.667 & 0.083 & 0.250 & 0.250 & 0.369 \\
$\omega_1$& 0 & 0.500 & 0.222 & 0.056 & 0.222 & 0.500 & 0.500 & 0.286 \\
$\omega_2$& 0 & 0.500 & 0.222 & 0.056 & 0.681 & 0.125 & 0.125 & 0.244 \\
$\omega_3$& 0 & 0 & 0.222 & 0.222 & 0.0139 & 0.125 & 0.125 & 0.101 \\
 \hline
\end{tabular}
\end{subtable}
\hfill
\begin{subtable}[t]{0.45\textwidth}
\begin{tabular}[t]{|c|ccccccc|c|}
\hline
\multicolumn{9}{|c|}{$\ket{\psi}=\ket{D_2}$}\\
\hline
&
$D_1$&
$S_1$&
$P_1$&
$F$&
$P_2$&
$S_2$&
$D_2$&
$\sum/7$
\\
\hline
 $\omega_0$&0.250 & 0.250 & 0.083 & 0.667 & 0.333 & 0 & 1.000 & 0.369 \\
 $\omega_1$&0.125 & 0.125 & 0.0139 & 0.222 & 0.222 & 0 & 0 & 0.101 \\
 $\omega_2$&0.125 & 0.125 & 0.681 & 0.056 & 0.222 & 0.500 & 0 & 0.244 \\
 $\omega_3$&0.500 & 0.500 & 0.222 & 0.056 & 0.222 & 0.500 & 0 & 0.286 \\
 \hline
\end{tabular}
\end{subtable}
\hfill

\bigskip

\hfill
\begin{subtable}[t]{0.45\textwidth}
\begin{tabular}{|c|ccccccc|c|}
\hline
\multicolumn{9}{|c|}{$\ket{\psi}=\ket{P_1}$}\\
\hline
&
$D_1$&
$S_1$&
$P_1$&
$F$&
$P_2$&
$S_2$&
$D_2$&
$\sum/7$
\\
\hline
 $\omega_0$& 0.333 & 0 & 1.000 & 0 & 0.250 & 0.750 & 0.083 & 0.345 \\
 $\omega_1$&0 & 0.167 & 0 & 0.167 & 0 & 0.167 & 0.167 & 0.095 \\
 $\omega_2$&0 & 0.167 & 0 & 0.167 & 0.375 & 0.042 & 0.375 & 0.161 \\
 $\omega_3$&0.667 & 0.667 & 0 & 0.667 & 0.375 & 0.042 & 0.375 & 0.399 \\
 \hline
\end{tabular}
\end{subtable}
\hfill
\begin{subtable}[t]{0.45\textwidth}
\begin{tabular}{|c|ccccccc|c|}
\hline
\multicolumn{9}{|c|}{$\ket{\psi}=\ket{P_2}$}\\
\hline
&
$D_1$&
$S_1$&
$P_1$&
$F$&
$P_2$&
$S_2$&
$D_2$&
$\sum/7$
\\
\hline
 $\omega_0$& 0.083 & 0.750 & 0.250 & 0 & 1.000 & 0 & 0.333 & 0.345 \\
 $\omega_1$& 0.375 & 0.083 & 0.375 & 0.667 & 0 & 0.667 & 0.667 & 0.399 \\
 $\omega_2$& 0.375 & 0.083 & 0.375 & 0.167 & 0 & 0.167 & 0 & 0.161 \\
 $\omega_3$& 0.167 & 0.167 & 0 & 0.167 & 0 & 0.167 & 0 & 0.095 \\
 \hline
\end{tabular}
\end{subtable}
\hfill

\bigskip

\hfill
\begin{subtable}[t]{0.45\textwidth}
\begin{tabular}[t]{|c|ccccccc|c|}
\hline
\multicolumn{9}{|c|}{$\ket{\psi}=\ket{S_1}$}\\
\hline
&
$D_1$&
$S_1$&
$P_1$&
$F$&
$P_2$&
$S_2$&
$D_2$&
$\sum/7$
\\
\hline
 $\omega_0$&0 & 1.000 & 0 & 0 & 0.750 & 0.250 & 0.250 & 0.321 \\
 $\omega_1$&0.500 & 0 & 0.500 & 0.500 & 0 & 0.500 & 0.500 & 0.286 \\
 $\omega_2$& 0.500 & 0 & 0.500 & 0.500 & 0.125 & 0.125 & 0.125 & 0.196 \\
 $\omega_3$&0 & 0 & 0 & 0 & 0.125 & 0.125 & 0.125 & 0.054 \\
 \hline
\end{tabular}
\end{subtable}
\hfill
\begin{subtable}[t]{0.45\textwidth}
\begin{tabular}[t]{|c|ccccccc|c|}
\hline
\multicolumn{9}{|c|}{$\ket{\psi}=\ket{S_2}$}\\
\hline
 &
$D_1$&
$S_1$&
$P_1$&
$F$&
$P_2$&
$S_2$&
$D_2$&
$\sum/7$
\\
\hline
 $\omega_0$&0.250 & 0.250 & 0.750 & 0 & 0 & 1.000 & 0 & 0.321 \\
 $\omega_1$&0.125 & 0.125 & 0.125 & 0 & 0 & 0 & 0 & 0.054 \\
 $\omega_2$&0.125 & 0.125 & 0.125 & 0.500 & 0.500 & 0 & 0.500 & 0.268 \\
 $\omega_3$&0.500 & 0.500 & 0 & 0.500 & 0.500 & 0 & 0.500 & 0.357 \\
 \hline
\end{tabular}
\end{subtable}
\hfill

\bigskip

\hfill
\begin{subtable}[t]{0.45\textwidth}
\begin{tabular}{|c|ccccccc|c|}
\hline
\multicolumn{9}{|c|}{$\ket{\psi}=\ket{F}$}\\
\hline
 &
$D_1$&
$S_1$&
$P_1$&
$F$&
$P_2$&
$S_2$&
$D_2$&
$\sum/7$
\\
\hline
 $\omega_0$& 0.667 & 0 & 0 & 1.000 & 0 & 0 & 0.667 & 0.333 \\
 $\omega_1$&0 & 0.333 & 0.333 & 0 & 0.333 & 0.333 & 0.333 & 0.238 \\
 $\omega_2$&0 & 0.333 & 0.333 & 0 & 0.333 & 0.333 & 0 & 0.190 \\
 $\omega_3$&0.333 & 0.333 & 0.333 & 0 & 0.333 & 0.333 & 0 & 0.238 \\
 \hline
\end{tabular}
\end{subtable}
\hfill
\begin{subtable}[t]{0.45\textwidth}
\begin{tabular}{|c|ccccccc|c|}
\hline
\multicolumn{9}{|c|}{$\ket{\psi}=\ket{\omega_1}$}\\
\hline
&
$D_1$&
$S_1$&
$P_1$&
$F$&
$P_2$&
$S_2$&
$D_2$&
$\sum/7$
\\
\hline
 $\omega_0$& 0.500 & 0.500 & 0.167 & 0.333 & 0.667 & 0 & 0 & 0.310 \\
 $\omega_1$&0.250 & 0.250 & 0.694 & 0.444 & 0.111 & 1.000 & 1.000 & 0.536 \\
 $\omega_2$&0.250 & 0.250 & 0.028 & 0.111 & 0.111 & 0 & 0 & 0.107 \\
 $\omega_3$&0 & 0 & 0.111 & 0.111 & 0.111 & 0 & 0 & 0.048 \\
 \hline
\end{tabular}
\end{subtable}
\hfill

\bigskip

\hfill
\begin{subtable}[t]{0.45\textwidth}
\begin{tabular}{|c|ccccccc|c|}
\hline
\multicolumn{9}{|c|}{$\ket{\psi}=\ket{\omega_2}$}\\
\hline
&
$D_1$&
$S_1$&
$P_1$&
$F$&
$P_2$&
$S_2$&
$D_2$&
$\sum/7$
\\
\hline
 $\omega_0$&0.500 & 0.500 & 0.167 & 0.333 & 0.167 & 0.500 & 0.500 & 0.381 \\
 $\omega_1$&0.250 & 0.250 & 0.0278 & 0.111 & 0.111 & 0 & 0 & 0.107 \\
 $\omega_2$&0.250 & 0.250 & 0.694 & 0.444 & 0.694 & 0.250 & 0.250 & 0.405 \\
 $\omega_3$&0 & 0 & 0.111 & 0.111 & 0.028 & 0.250 & 0.250 & 0.107 \\
 \hline
\end{tabular}
\end{subtable}
\hfill
\begin{subtable}[t]{0.45\textwidth}
\begin{tabular}{|c|ccccccc|c|}
\hline
\multicolumn{9}{|c|}{$\ket{\psi}=\ket{\omega_3}$}\\
\hline
&
$D_1$&
$S_1$&
$P_1$&
$F$&
$P_2$&
$S_2$&
$D_2$&
$\sum/7$
\\
\hline
 $\omega_0$&0 & 0 & 0.667 & 0.333 & 0.167 & 0.500 & 0.500 & 0.310 \\
 $\omega_1$&0 & 0 & 0.111 & 0.111 & 0.111 & 0 & 0 & 0.048 \\
 $\omega_2$&0 & 0 & 0.111 & 0.111 & 0.028 & 0.250 & 0.250 & 0.107 \\
 $\omega_3$&1.000 & 1.000 & 0.111 & 0.444 & 0.694 & 0.250 & 0.250 & 0.536 \\
 \hline
\end{tabular}
\end{subtable}
\hfill

\bigskip

\hfill
\begin{subtable}[t]{0.45\textwidth}
\begin{tabular}{|c|ccccccc|c|}
\hline
\multicolumn{9}{|c|}{ $\ket{\psi}=(\ket{1}-3\ket{2}+\ket{3})/\sqrt{11}$}\\
\hline
&
$D_1$&
$S_1$&
$P_1$&
$F$&
$P_2$&
$S_2$&
$D_2$&
$\sum/7$
\\
\hline
 $\omega_0$& 0.727 & 0.182 & 0.545 & 0.273 & 0.545 & 0.182 & 0 & 0.351 \\
 $\omega_1$ & 0.091 & 0.091 & 0.091 &  0.364 & 0 & 0 & 0.091 & 0.104 \\
 $\omega_2$& 0.091 & 0.364 & 0 & 0 & 0.364 & 0 & 0.091 & 0.130 \\
 $\omega_3$& 0.091 & 0.364 & 0.364 & 0.364 & 0.091 & 0.818 & 0.818 & 0.416 \\
 \hline
\end{tabular}
\end{subtable}
\hfill
\begin{subtable}[t]{0.45\textwidth}
\begin{tabular}{|c|ccccccc|c|}
\hline
\multicolumn{9}{|c|}{ $\ket{\psi}=\ket{N_F} = (\ket{1}+\ket{2}+\ket{3})/\sqrt{3}$}\\
\hline
&
$D_1$&
$S_1$&
$P_1$&
$F$&
$P_2$&
$S_2$&
$D_2$&
$\sum/7$
\\
\hline
 $\omega_0$& 0 & 0.667 & 0.222 & 0.111 & 0.222 & 0.667 & 0 & 0.270\\
 $\omega_1$ & 0.333 & 0.333 & 0.037 & 0.148 & 0.593 & 0 & 0.333 & 0.254\\
 $\omega_2$&  0.333 & 0 & 0.148 & 0.593 & 0.148 & 0 & 0.333 & 0.032\\
 $\omega_3$&  0.333 & 0 & 0.593 & 0.148 & 0.037 & 0.333 & 0.333 & 0.254\\
 \hline
\end{tabular}
\end{subtable}
\hfill

\bigskip

\hfill
\begin{subtable}[t]{0.45\textwidth}
\begin{tabular}{|c|ccccccc|c|}
\hline
\multicolumn{9}{|c|}{ $\ket{\psi}=\ket{N_{S1}} = (2\ket{1}+\ket{2})/\sqrt{5}$}\\
\hline
&
$D_1$&
$S_1$&
$P_1$&
$F$&
$P_2$&
$S_2$&
$D_2$&
$\sum/7$
\\
\hline
 $\omega_0$& 0.100 & 0.100 & 0.300 & 0.600 & 0 & 0.400 & 0.400 & 0.271\\
 $\omega_1$ & 0.800 & 0.800 & 0.200 & 0.200 & 0.800 & 0.200 & 0.200 & 0.457\\
 $\omega_2$&  0.050 & 0.050 & 0.050 & 0.200 & 0 & 0.200 & 0.200 & 0.107\\
 $\omega_3$&  0.050 & 0.050 & 0.450 & 0 & 0.200 & 0.200 & 0.200 & 0.164\\
 \hline
\end{tabular}
\end{subtable}
\hfill
\begin{subtable}[t]{0.45\textwidth}
\begin{tabular}{|c|ccccccc|c|}
\hline
\multicolumn{9}{|c|}{ $\ket{\psi}=\ket{N_{X}} = (2\ket{1}+2\ket{2}+\ket{3})/3$}\\
\hline
&
$D_1$&
$S_1$&
$P_1$&
$F$&
$P_2$&
$S_2$&
$D_2$&
$\sum/7$
\\
\hline
 $\omega_0$& 0.055 & 0.500 & 0.167 & 0.333 & 0.167 & 0.500 & 0.055 & 0.254\\
 $\omega_1$ & 0.444 & 0.444 & 0.111 & 0.111 & 0.694 & 0.111 & 0.250 & 0.298\\
 $\omega_2$& 0.250 & 0.111 & 0.111 & 0.444 & 0.111 & 0.111 & 0.250 & 0.151 \\
 $\omega_3$& 0.250 & 0.028 & 0.694 & 0.111 & 0.111 & 0.444 & 0.444 & 0.298 \\
 \hline
\end{tabular}
\end{subtable}
\hfill

\bigskip

\hfill
\begin{subtable}[t]{0.45\textwidth}
\begin{tabular}{|c|ccccccc|c|}
\hline
\multicolumn{9}{|c|}{ $\ket{\psi}=\ket{N_{1}} = (\ket{1}+3\ket{2}+\ket{3})/\sqrt{11}$}\\
\hline
&
$D_1$&
$S_1$&
$P_1$&
$F$&
$P_2$&
$S_2$&
$D_2$&
$\sum/7$
\\
\hline
 $\omega_0$& 0.182 & 0.727 & 0 & 0.273 & 0.545 & 0.182 & 0 & 0.273\\
 $\omega_1$ & 0.091 & 0.091 & 0.091 & 0 & 0.364 & 0 & 0.091 & 0.104\\
 $\omega_2$& 0.364 & 0.091 & 0.091 & 0.364 & 0 & 0 & 0.091 & 0.143\\
 $\omega_3$& 0.364 & 0.091 & 0.818 & 0.364 & 0.091 & 0.818 & 0.818 & 0.481 \\
 \hline
\end{tabular}
\end{subtable}
\hfill
\begin{subtable}[t]{0.45\textwidth}
\begin{tabular}{|c|ccccccc|c|}
\hline
\multicolumn{9}{|c|}{ $\ket{\psi}= (\ket{1}+ \ket{2})/\sqrt{2}$}\\
\hline
&
$D_1$&
$S_1$&
$P_1$&
$F$&
$P_2$&
$S_2$&
$D_2$&
$\sum/7$
\\
\hline
 $\omega_0$& 0.250 & 0.250 & 0.083 & 0.667 & 0.083 & 0.250 & 0.250 & 0.262\\
 $\omega_1$& 0.500 & 0.500 & 0.222 & 0.056 & 0.681 & 0.125 & 0.125 & 0.315\\
 $\omega_3$& 0.125 & 0.125 & 0.014 & 0.222 & 0.014 & 0.125 & 0.125 & 0.107\\
 $\omega_3$& 0.125 & 0.125 & 0.681 & 0.056 & 0.222 & 0.500 & 0.500 & 0.315\\
 \hline
\end{tabular}
\end{subtable}
\hfill
\end{table}

\end{widetext}

\end{appendix}

\end{document}